\documentclass[iop]{emulateapj}
\usepackage{apjfonts}
\usepackage{graphicx}
\usepackage{amsmath}
\usepackage{hyperref}
\hypersetup{
    colorlinks=true,       
    linkcolor=blue,        
    citecolor=blue,        
    urlcolor=blue           
}

\graphicspath{{./figs/}}

\newcommand{\Msun}{\rm M_{\odot}}

\def\Omm{{\Omega_m}}

\def\Omb{{\Omega_b}}
\def\Oml{{\Omega_{\Lambda}}}

\def\beq{\begin{equation}}
\def\eeq{\end{equation}}

\shorttitle{}
\shortauthors{}

\begin{document}

\title{The effects of local primordial non-Gaussianity on the formation and evolution of galaxies}

\author
{
Xinghai Zhao \altaffilmark{1,3},
Yuexing Li \altaffilmark{1,3},
Sarah Shandera \altaffilmark{2,3} and
Donghui Jeong \altaffilmark{4}
}

\affil{$^{1}$ Department of Astronomy \& Astrophysics, The Pennsylvania State University, 525 Davey Lab, University Park, PA 16802, USA}
\affil{$^{2}$ Department of Physics, The Pennsylvania State University, 104 Davey Lab, University Park, PA 16802, USA}
\affil{$^{3}$Institute for Gravitation and the Cosmos, The Pennsylvania State University, 104 Davey Lab, University Park, PA 16802, USA}
\affil{$^{4}$ Department of Physics and Astronomy, Johns Hopkins University, 3400 N. Charles Street, Baltimore, Maryland 21210, USA}

\email{xuz21@psu.edu}

\begin{abstract}

Thanks to the rapid progress in precision cosmology in the last few years, we now have access to physical observables that may constrain the theory of inflation through the non-Gaussianity (NG) signatures in the cosmic microwave background radiation and the distribution of large-scale structure. Numerical modeling of the NG signals from different inflation models is essential to correctly interpret current and near future data from large-scale structure surveys. In this study, we use high-resolution cosmological hydrodynamical simulations to investigate the effects of primordial NG on the formation and evolution of galaxies from the cosmic dawn to the present day. Focusing on the local type primordial NG, we find that it may affect the formation history of stars and black holes in galaxies, and their  distribution. Compared to the Gaussian case, large non-Gaussian potential with $f_{NL} \gtrsim 10^3$ leads to earlier collapse of the first structures, more massive galaxies especially at high redshifts, stronger clustering of galaxies, and higher halo bias. However, for smaller NG with $f_{NL} \lesssim 10^2$, the effect is significantly weaker. Observations of the distribution and properties of high-redshift, rare objects such as the first galaxies and quasars may provide further constraints on the primordial NG.

\end{abstract}


\section{Introduction}
\label{s1}

The recent developments in the Cosmic Microwave Background (CMB) physics have transformed our understanding of the origin of the Universe and suggested that the density at high redshift ($z \sim 1100$) can be described by a uniform background plus some small fluctuation of $ \sim 10^{-5}$ \citep{Bennett:2012, Planck2013}. These primordial density fluctuations then evolve into the cosmic structures that we observe today. The leading theory on the origin of such initial density fluctuations is that they were seeded by quantum fluctuations \citep{Mukhanov:1981,Hawking:1982,Starobinsky:1982,Guth:1982,Bardeen:1983} that were generated during the epoch of inflation \citep{Starobinsky:1980, Guth:1982, Albrecht:1982, Linde:1982}. 

Minimal single-field, slow-roll inflation models predict that these fluctuations are Gaussian to at least one part in $10^{6}$ \citep{Acquaviva:2003,Maldacena:2003,Bartolo:2004}. However, there is a rich variety of well-motivated inflation models that produce initial density fluctuations consistent with observations but with significant signatures of non-Gaussianity (NG, \citealt{Bartolo:2004}). Studying the detailed characteristics of primordial NG, such as three-point correlation function or bispectrum, enables us to distinguish between qualitatively different physics of the inflationary era. For example, the local type primordial NG we consider here is a clear signature of inflation scenarios with more than one scalar field \citep{Creminelli:2004}.

The local type primordial NG can be expressed by the gauge-invariant gravitational potential as follows \citep{Salopek:1990,Komatsu:2001}, 
\begin{equation}
\Phi({\bf{x}})=\Phi_G({\bf{x}})+f_{NL}\left[\Phi_G^2({\bf{x}})-\langle\Phi_G^2({\bf{x}})\rangle\right],
\label{eq_fnl_local}
\end{equation}
where $\Phi_G$ describes the linear Gaussian random field, and $f_{NL}$ is a dimensionless parameter that describes the magnitude of NG. Throughout this study, $f_{NL}$ represents the value of the local type primordial NG.

Constraints on the value of $f_{NL}$ have been obtained from the CMB measurements and surveys of large-scale structure. The current data from the Wilkinson Microwave Anisotropy Probe (WMAP) suggests that $-3<f_{NL}<77$ (95\% CL from WMAP9 bispectrum data, \citealt{Bennett:2012}), while the new measurement from the Planck satellite suggests that $f_{NL}=2.7 \pm 5.8$ \citep{Planck2013}. On the other hand, the abundance and clustering of the large-scale structure in the universe provide several promising tools to constrain primordial NG \citep{Sefusatti:2007,Verde:2010}. Limits on any scale-dependence in the large scale bias of quasars and galaxies have provided a constraint on local NG comparable to that from the CMB (e.g., $-29<f_{NL}<70$ (95\% CL, SDSS LRG and quasar data) \citep{Slosar:2008}, $5<f_{NL}<84$ (95\% CL, NVSS radio source and SDSS quasar and LRG data) \citep{Xia:2011}, $-45<f_{NL}<195$ (95\% CL, SDSS-III BOSS CMASS data) \citep{Ross:2013} and $-37<f_{NL}<25$ (95\% CL) \cite{Giannantonio:2013}). Most importantly, future surveys (e.g., LSST, Euclid, DES and eROSITA) should offer a much tighter constraint on primordial NG ($f_{NL}\sim \mathcal{O}(1)$) \citep{Carbone:2008,Carbone:2010,Giannantonio:2012,Amendola:2012,Cunha:2010,Pillepich:2012}. This improvement over the Planck results would have important implications for inflation theory. Given how important NG studies are for understanding the primordial universe, this analysis will certainly be a key component of large scale structure surveys for at least the next decade, regardless of the bounds Planck reports earlier this year. In addition, some large-scale structure observations can study the primordial NG on the scales much smaller than those probed by the CMB. This is important for studying any scale-dependence of the NG \citep{LoVerde:2008,Sefusatti:2009,Shandera:2011,Becker:2012,Becker:2012a}. In short, primordial NG in large-scale structure is an indispensable tool, complementary to the CMB, for studying the origin of the density inhomogeneities \citep{Verde:2010}.

Two aspects of the large-scale structure that may be used to study NG are the abundance of high redshift structures \citep{Matarrese:2000,LoVerde:2008,Grossi:2009,Pillepich:2010,Maggiore:2010,Pillepich:2012} or voids \citep{Kamionkowski:2009,DAmico:2011}, and the halo bias \citep{Dalal:2008,Matarrese:2008,Afshordi:2008,Giannantonio:2010}. The abundance of objects or voids is insensitive to the shape of the bispectrum and so it can put useful constraints on the magnitude of NG of any type \citep{Carbone:2008,Carbone:2010,Giannantonio:2012,Amendola:2012}. The halo bias is sensitive only to the squeezed limit of the bispectrum (that is, any correlation between two short wavelength modes and a long wavelength mode, which is where the local type of NG has a strong signal). The characteristic properties of abundance of cosmic structures and of the halo bias have been intensively studied with both analytical and numerical approaches in the last decade. Substantial progress has been made on the study of the halo mass function \citep{Matarrese:2000,Dalal:2008,LoVerde:2008,Grossi:2009,Wagner:2010,LoVerde:2011} and the halo bias \citep{Dalal:2008,Matarrese:2008,Afshordi:2008,Shandera:2011,Desjacques:2011,Desjacques:2011a,Wagner:2012,Ferraro:2012,Schmidt:2012,Scoccimarro:2012,DAloisio:2013,Desjacques:2013}. For dark matter only simulations, both analytical and numerical approaches have been shown to be consistent with each other reasonably well (e.g. \cite{Grossi:2009,Desjacques:2009,Desjacques:2010a,Scoccimarro:2012}).  In addition, it is becoming clear how to utilize those features to constrain the primordial NG from large-scale structure surveys (see \citealt{Verde:2010} for a review).

However, until very recently all non-Gaussian simulations contained only the dark matter component and therefore have limited power in predicting NG signatures for comparison with large-scale structure surveys. In order to study effects of the primordial NG on the rich ``visible'' structures (galaxies, stars, black holes etc.) in the observable Universe, the baryonic component and its complex physics must be included in the realistic simulations. This is an absolutely crucial step toward using the data to constrain the primordial physics. For example, although the large-scale bias has been used to constrain NG \citep{Slosar:2008, Xia:2011,Ross:2013}, it is not clear whether the parameter constrained is the same $f_{NL}$ that the CMB constrains, and how the bias of visible galaxies is related to the NG of the primordial density field. 

Recently, non-Gaussian simulations with baryonic components that incorporate hydrodynamical and chemical processes have been performed by \cite{Maio:2011}, and these authors have studied the effects of primordial NG on primordial gas properties and star formation \citep{Maio:2011a, Maio:2012}. It was suggested by these studies that for a moderately non-Gaussian scenario (e.g. $0 \le f_{NL} \le 100$), the NG effects are insignificant except at the very early stage of the cosmic structure formation ($z \ge 10$). For the extreme case ($f_{NL}=1000$), the effects may be significant but this has been ruled out by the current constraints from observations \citep{Komatsu:2011,Slosar:2008,Xia:2011,Ross:2013, Giannantonio:2013, Planck2013}. The possibility of using long gamma-ray bursts (GRBs) at high redshift to constrain the primordial NG has been studied by \cite{Maio:2012a}. However, their utility is limited by the small sample of such GRB currently available, and GRBs have not yielded any comparable constraints as other observations \citep{Maio:2012a}.

In this paper, we extend the previous studies to investigate the effects of NG on the formation and evolution of galaxies by using cosmological hydrodynamical simulations with higher resolutions and additional physical processes such as black hole accretion and feedback. We focus on the two important aspects of structure formation that may give constraints of NG, the abundance and bias of galaxies. This paper is organized as follows: In Sec. \ref{s2}, we describe the NG model and a suite of cosmological simulations with Gaussian and NG initial conditions. In Sec. \ref{s3}, we present the results of galaxy properties such as the mass functions and the cosmic formation history of stars and black holes, and galaxy bias, from different NG simulations. We discuss the implications and limitations of our models, and summarize in Sec. \ref{s4}.

\section{Methodology}
\label{s2}

\subsection{Initial Conditions}
\label{ss21}

In this study, we perform a set of cosmological simulations from both Gaussian and non-Gaussian initial conditions. For the NG simulations, we focus on the local type because it is well physically motivated, intensively studied and it is very sensitive to the quantities from the large-scale surveys and thus tightly constrained. 

The initial conditions of the simulations are set up with the code developed by \cite{Scoccimarro:2012}. The code can be used to generate several types of non-Gaussian initial conditions, but in this study we use the local type initial conditions with $f_{NL}=0, 100, 1000$. We only study $f_{NL}\geq 0$ in this work since recent observations all indicate a central value of $f_{NL}$ that is positive \citep{Bennett:2012,Komatsu:2011,Slosar:2008,Xia:2011,Ross:2013}. The $f_{NL}$ values in the initial conditions (0, 100, 1000) are in the CMB convention for the primordial potential. In the theoretical fittings using the data from the large-scale structure (LSS) presented later in this paper, the LSS convention is used instead due to the growth of the potential in the structure formation process, $f_{NL}^{LSS}=1.3f_{NL}^{CMB}$ \citep{Matarrese:2000,Dalal:2008,LoVerde:2008,Grossi:2009,Verde:2010} in these cases.

The displacements and velocities of the particles in the initial conditions are calculated by the second order Lagrangian perturbation theory (2LPT) \citep{Scoccimarro:1998,Jenkins:2010,Hahn:2011,Scoccimarro:2012}. It has been shown \citep{Munshi:1994,Scoccimarro:1998} that the Zel'Dovich approximation \citep{Zeldovich:1970} or the first order Lagrangian perturbation theory (1LPT) can substantially underestimate higher order moments of the probability distribution function (PDF) of the density field at high redshift because it takes a non-negligible period of time to make a transient to disappear so that the simulations reflect the correct statistics of the density and velocity fields \citep{Scoccimarro:1998}. 2LPT is much less affected by this problem and most importantly, it can correctly reproduce the skewness of the density field \citep{Munshi:1994} that develops even from Gaussian initial conditions. 
Therefore, when 2LPT is used, the simulations can start at much lower redshifts than the ones that are set up with 1LPT while still achieving comparable accuracy \citep{Jenkins:2010,Scoccimarro:2012}. The starting redshift $z=99$ of our simulations should be sufficient for the investigation of the effects of the primordial NG discussed in this study.

All of the settings in the Gaussian and NG simulations are identical except the degree of NG set in the initial conditions. For each type of the simulations, we perform both dark matter-only and hydrodynamical simulations that contain both dark matter and baryonic matter (gas). The initial conditions are set up in a box of 100 Mpc/h in comoving coordinates with a periodic boundary condition. Each of the dark matter and gas components is represented by $512^3$ particles, respectively, which yields a gravitational softening length $\epsilon=$ 5 kpc/h, and a particle mass of $4.65 \times 10^7\Msun$/h for the gas and stars, and $4.65 \times 10^8\Msun$/h for the dark matter. The mass resolutions improve over those of \cite{Maio:2011} by a factor of about 5, which allow us to study galaxy mass functions at high redshifts. 

The cosmological parameters adopted in the simulations are:  $\Omm=0.27$, $\Omb=0.045$, $\Oml=0.73$, $\sigma_8=0.8$, $n_s=1.0$ and $H_0=100$h km/s/Mpc with $\rm h=0.7$ in a $\Lambda CDM$ cosmology that is consistent with the recent WMAP7 results \citep{Komatsu:2011}. The simulations start from the initial conditions at $z=99$ and trace the structure formation and evolution processes until current time at $z=0$.

\subsection{Cosmological Simulations}
\label{ss22}

The simulations in this study are performed using the N-body TreePM SPH code GADGET-3 \citep{Springel:2008} that is an improved version of the widely used simulation code GADGET-2 \citep{Springel:2001a,Springel:2005,Springel:2005a}. For gravity, GADGET-3 utilizes the TreePM method \citep{Xu:1995} that calculates the short-range gravitational force with the tree method \citep{Barnes:1986} and the long-range gravitational force with the particle-mesh method \citep{Hockney:1981}. Thus the code can calculate both the long-range and short-range gravitational force efficiently while achieving high local resolution in the simulations. 

The hydrodynamical simulations include important baryonic physics such as gas cooling and heating, star formation, supernova feedback, and black hole accretion and feedback. For the hydrodynamics, GADGET-3 uses the smoothed-particle hydrodynamics (SPH) (see \cite{Monaghan:1992} for a review) formalism with an entropy-conserving formulation presented in \cite{Springel:2002} that can conserve the energy and entropy simultaneously with adapting particle smoothing lengths. The gas cooling process is modeled with the method presented in \cite{Katz:1996} using two-body processes that include excitation, ionization, recombination and emission. The gas heating process includes the UV background model presented in \cite{Faucher-Giguere:2009} in which the cosmic reionization is assumed to be finished by $z=6$. 

The star formation process is modeled with a hybrid multi-phase interstellar medium (ISM) model \citep{Springel:2003,Springel:2003a} in which the star formation rate is calculated with the Schmidt-Kennicutt law \citep{Schmidt:1959,Kennicutt:1998}. The supernova feedback implemented in the code contains the thermal heating and cloud evaporation from supernova thermal feedback and the galactic wind and outflow from supernova kinetic feedback \citep{Springel:2003,Springel:2003a}. The black hole accretion and feedback are modeled using the model presented in \cite{Springel:2005b} and \cite{Di-Matteo:2005}. In this model the black holes are represented by collisionless sink particles that can accrete mass from surrounding gas particles with a rate estimated with the Bondi accretion model \citep{Bondi:1952}. The black holes can also grow by merging with other nearby black holes. The black hole feedback is taken as thermal feedback in which a fraction of the black hole radiation energy deposits isotropically to the surrounding gas. For the origin of the black hole particles, the initial black hole seeds with mass of $10^5\Msun$/h in the simulations are planted at the center of the galaxies of which total mass reaches $10^{10}\Msun$/h if they haven't contained a black hole yet as in \cite{Di-Matteo:2008}. This self-regulated black hole growth model has been shown to produce results that agree with many observed properties of the local galaxies and distant quasars \citep{Springel:2005b,Springel:2005a,Di-Matteo:2005,Di-Matteo:2008,Hopkins:2006,Li:2007,Zhu:2012}.

In order to find structures formed on different scales in the simulations, a two-step structure finding procedure is performed on-the-fly for each snapshots along the simulation. First, a friends-of-firends (FOF) algorithm is used to identify the particle groups in the snapshots by linking the dark matter particles with particle separations less than 0.2 times of the mean particle spacing in the simulations. The gas, star and black hole particles are then linked to the identified dark matter particle groups using the same algorithm. Second, the SUBFIND algorithm \citep{Springel:2001,Dolag:2009} is used to find gravitationally bound physical substructures in the groups. The SUBFIND first finds the local overdensities using the SPH kernel interpolation. It then iteratively removes the gravitationally unbound particles. If the final clumps have more than 20 particles, we treat them as physically bound substructures.

\section{Results}
\label{s3}

\begin{figure*}[h]
\begin{center}
\includegraphics[trim=0cm 0cm 0cm 0cm, clip=true, angle=0, width=2.125in]{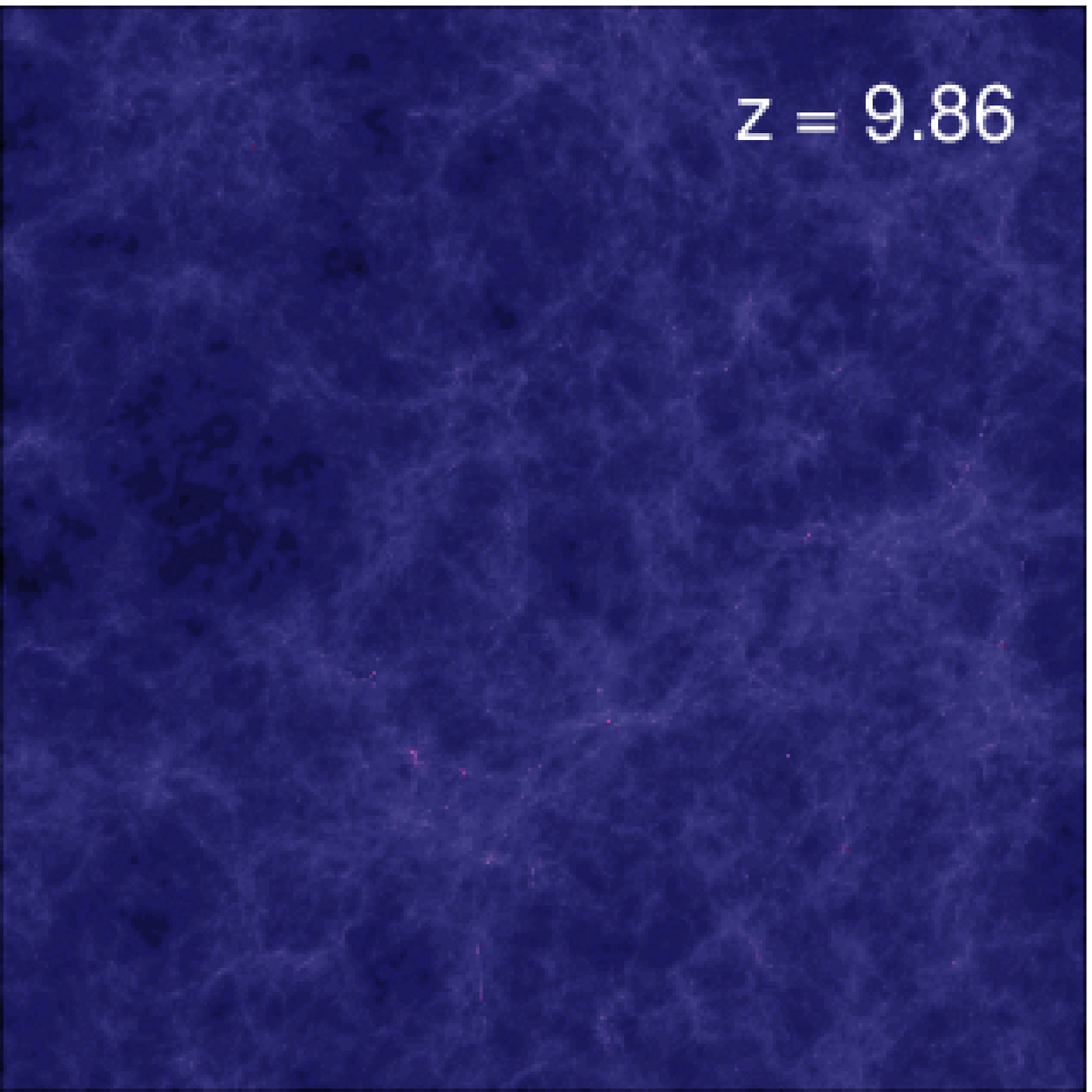}
\includegraphics[trim=0cm 0cm 0cm 0cm, clip=true, angle=0, width=2.125in]{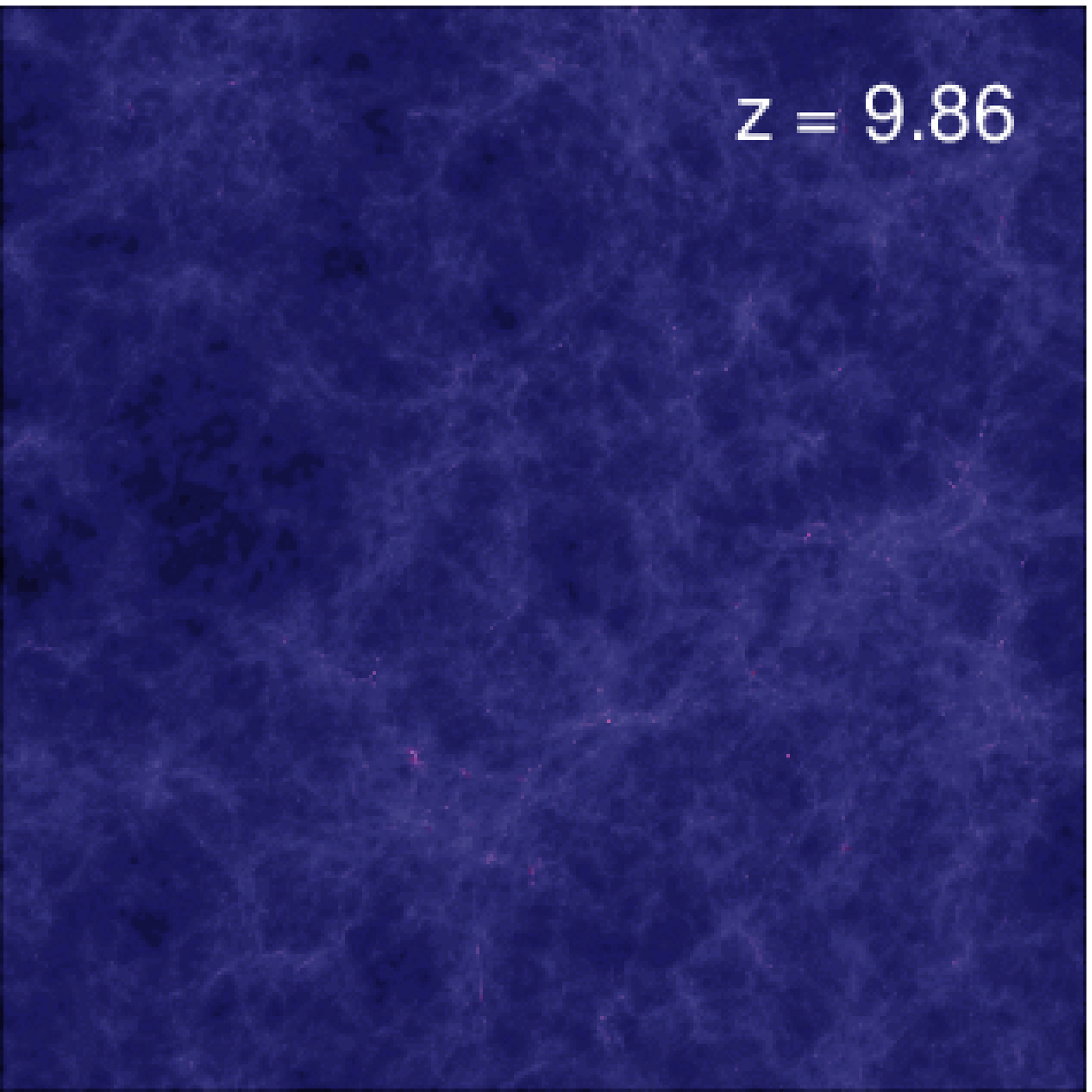}
\includegraphics[trim=0cm 0cm 0cm 0cm, clip=true, angle=0, width=2.125in]{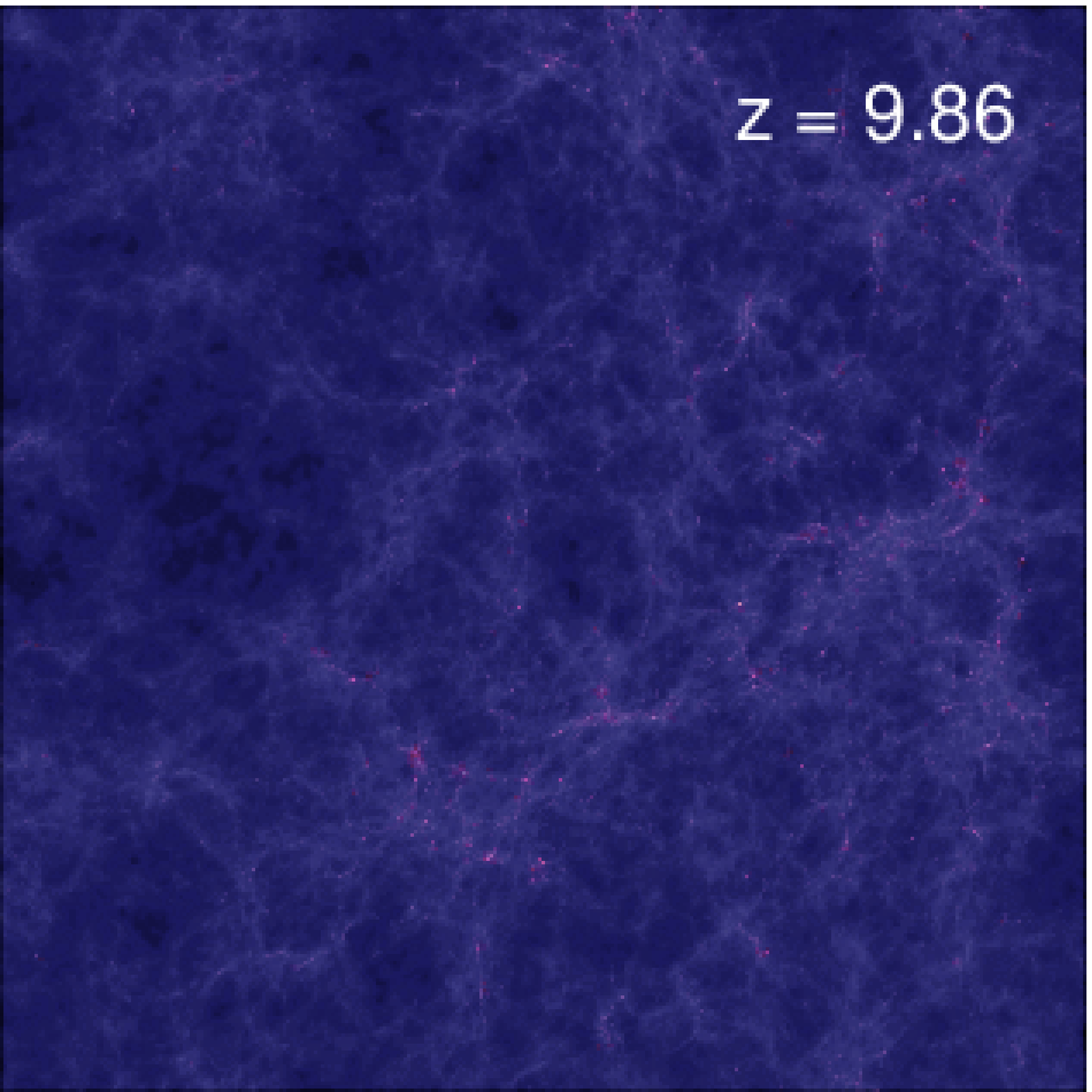} \\
\includegraphics[trim=0cm 0cm 0cm 0cm, clip=true, angle=0, width=2.125in]{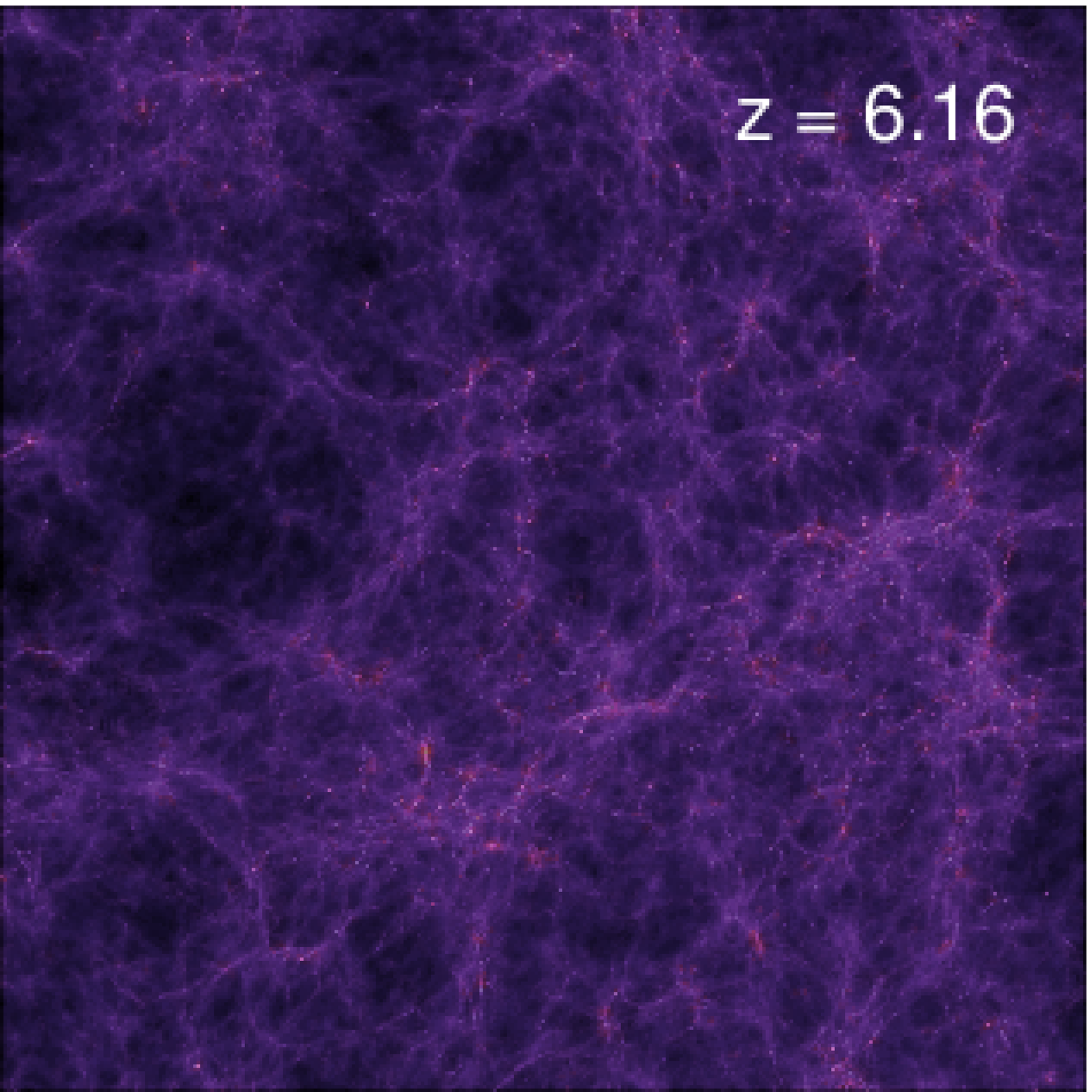}
\includegraphics[trim=0cm 0cm 0cm 0cm, clip=true, angle=0, width=2.125in]{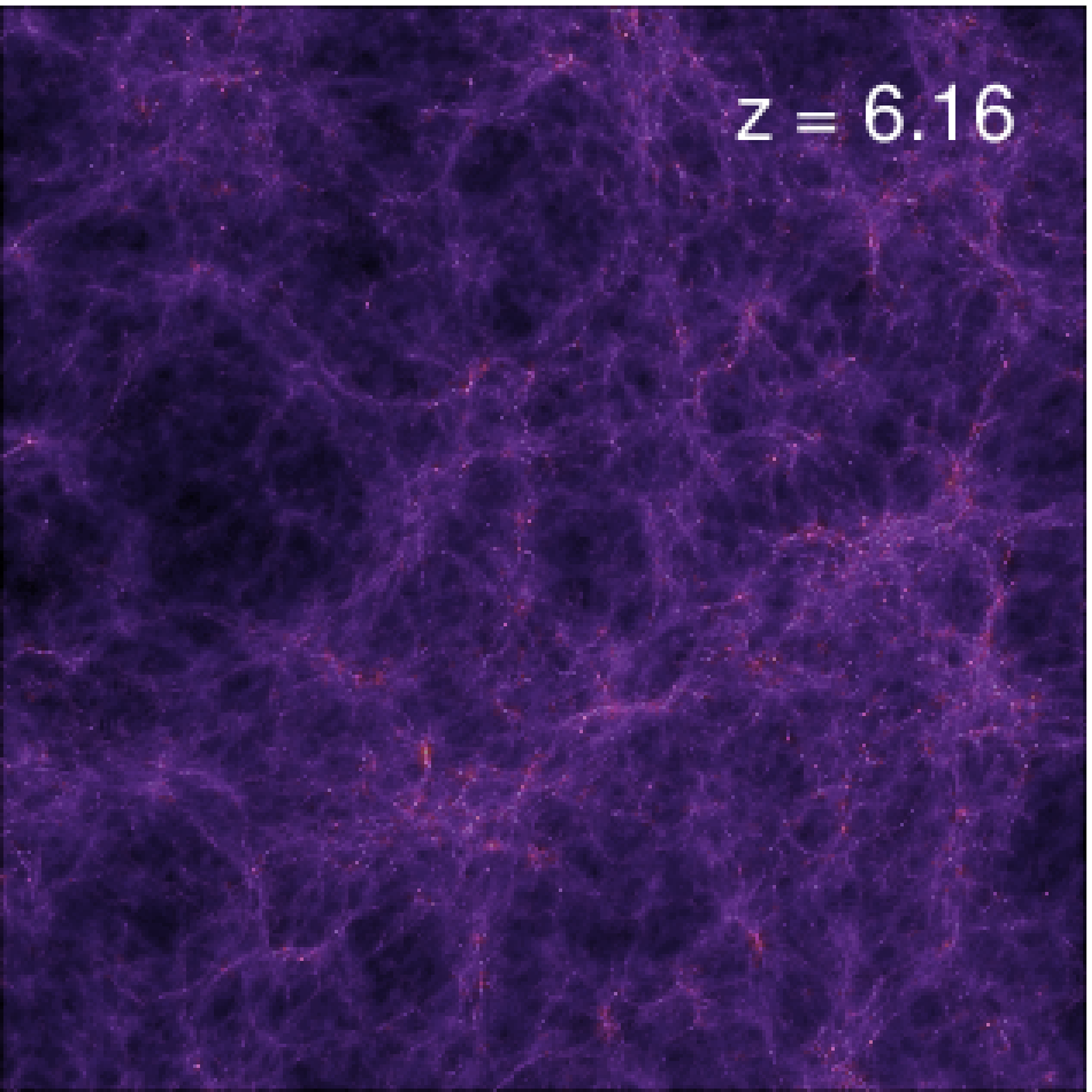}
\includegraphics[trim=0cm 0cm 0cm 0cm, clip=true, angle=0, width=2.125in]{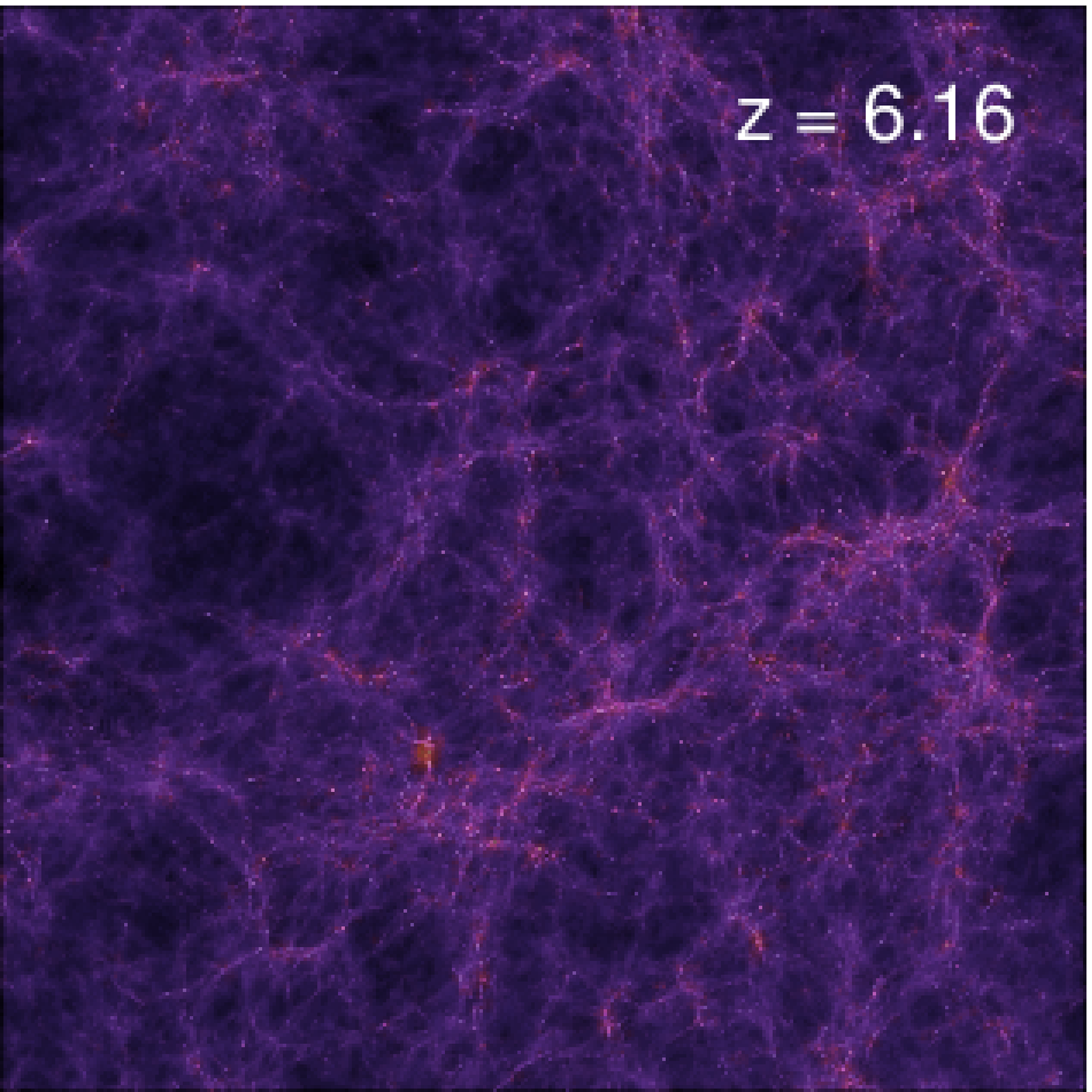} \\
\includegraphics[trim=0cm 0cm 0cm 0cm, clip=true, angle=0, width=2.125in]{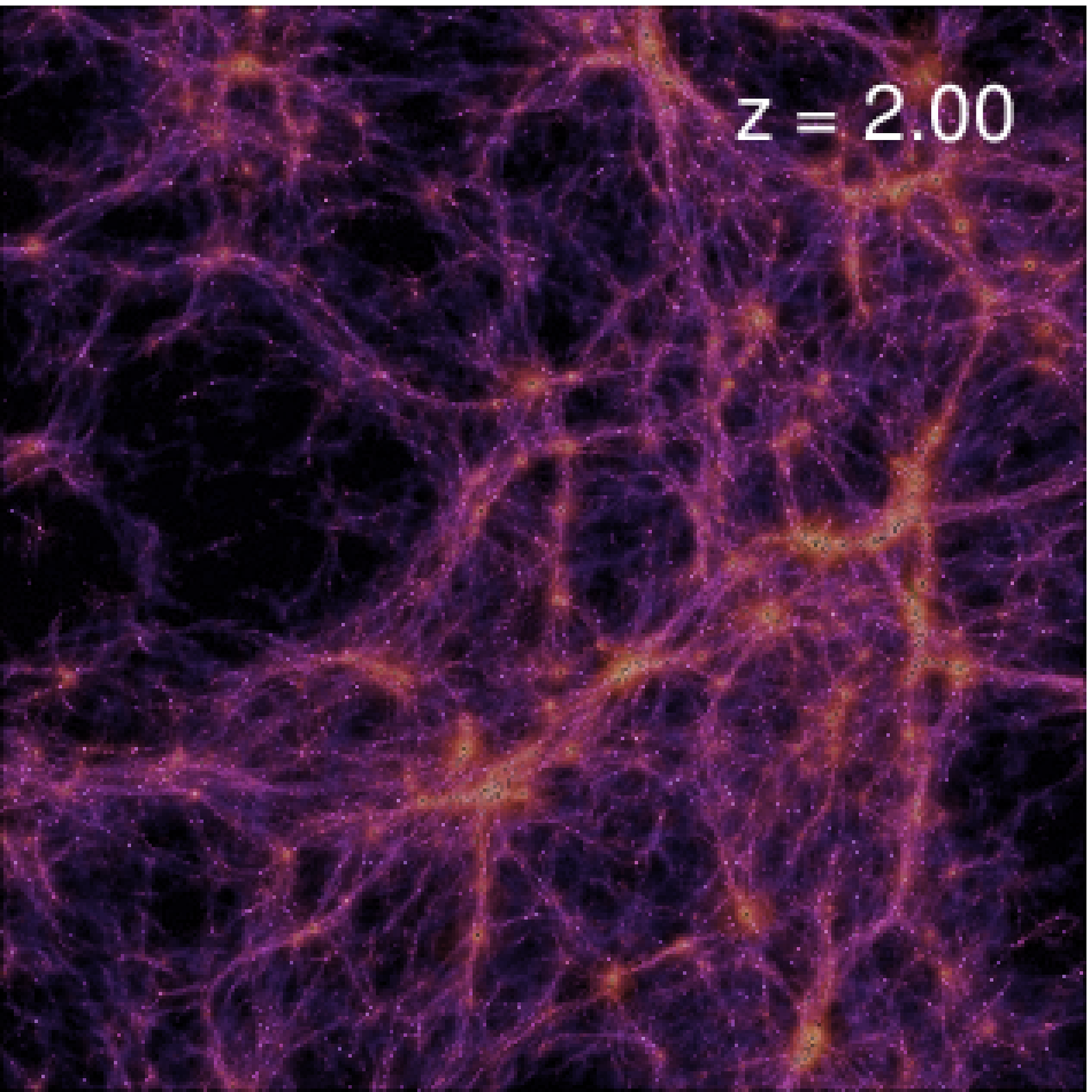}
\includegraphics[trim=0cm 0cm 0cm 0cm, clip=true, angle=0, width=2.125in]{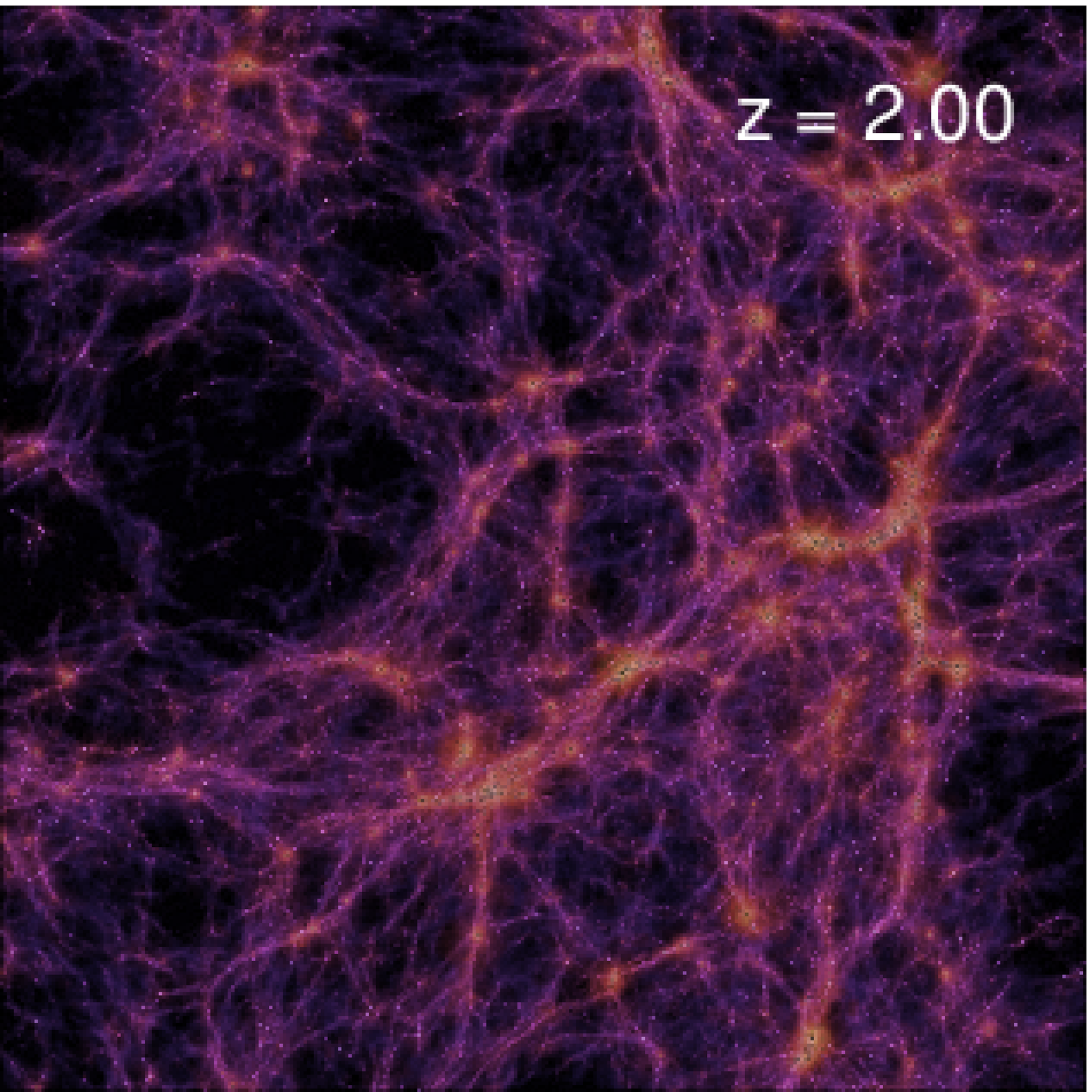}
\includegraphics[trim=0cm 0cm 0cm 0cm, clip=true, angle=0, width=2.125in]{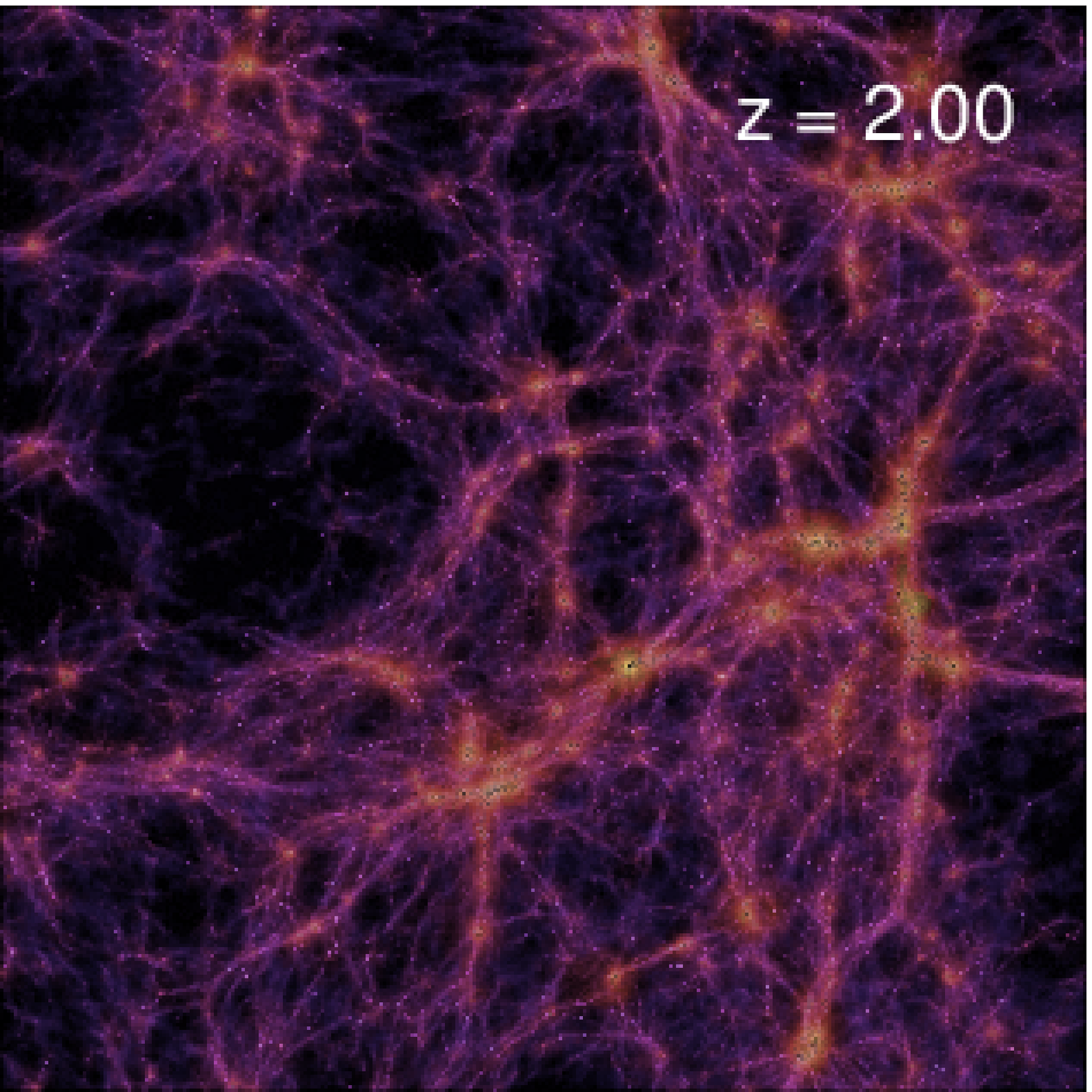} \\
\includegraphics[trim=0cm 0cm 0cm 0cm, clip=true, angle=0, width=2.125in]{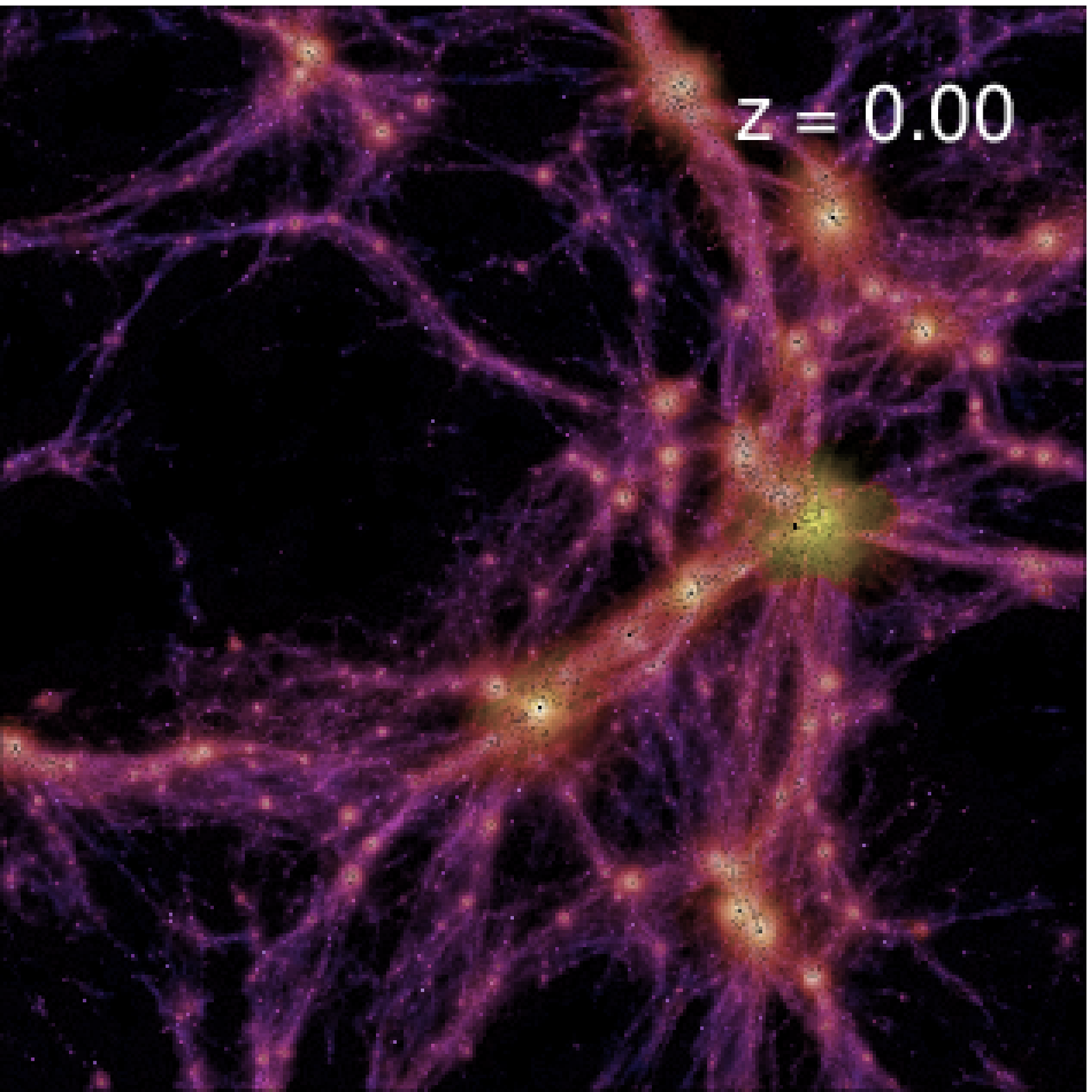}
\includegraphics[trim=0cm 0cm 0cm 0cm, clip=true, angle=0, width=2.125in]{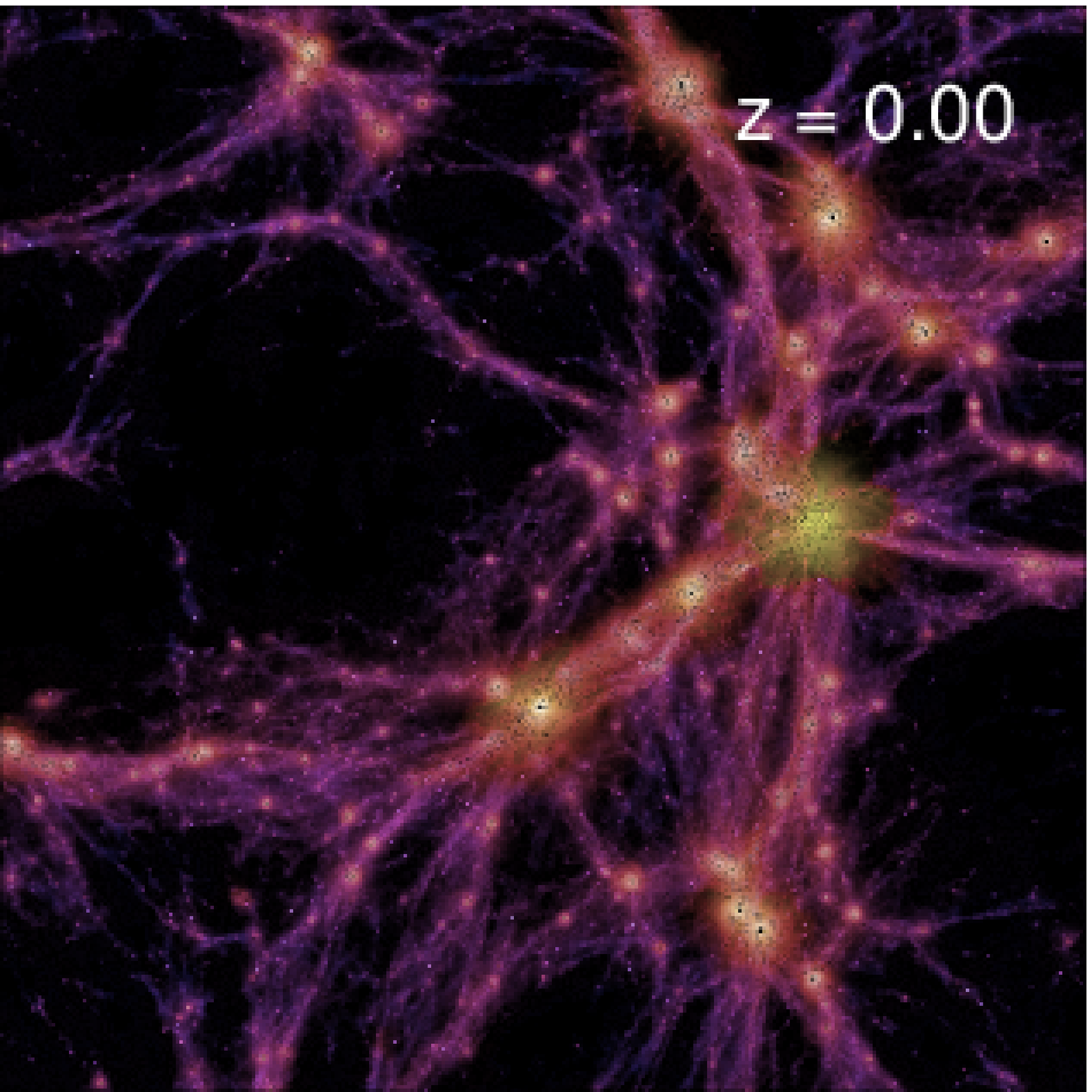}
\includegraphics[trim=0cm 0cm 0cm 0cm, clip=true, angle=0, width=2.125in]{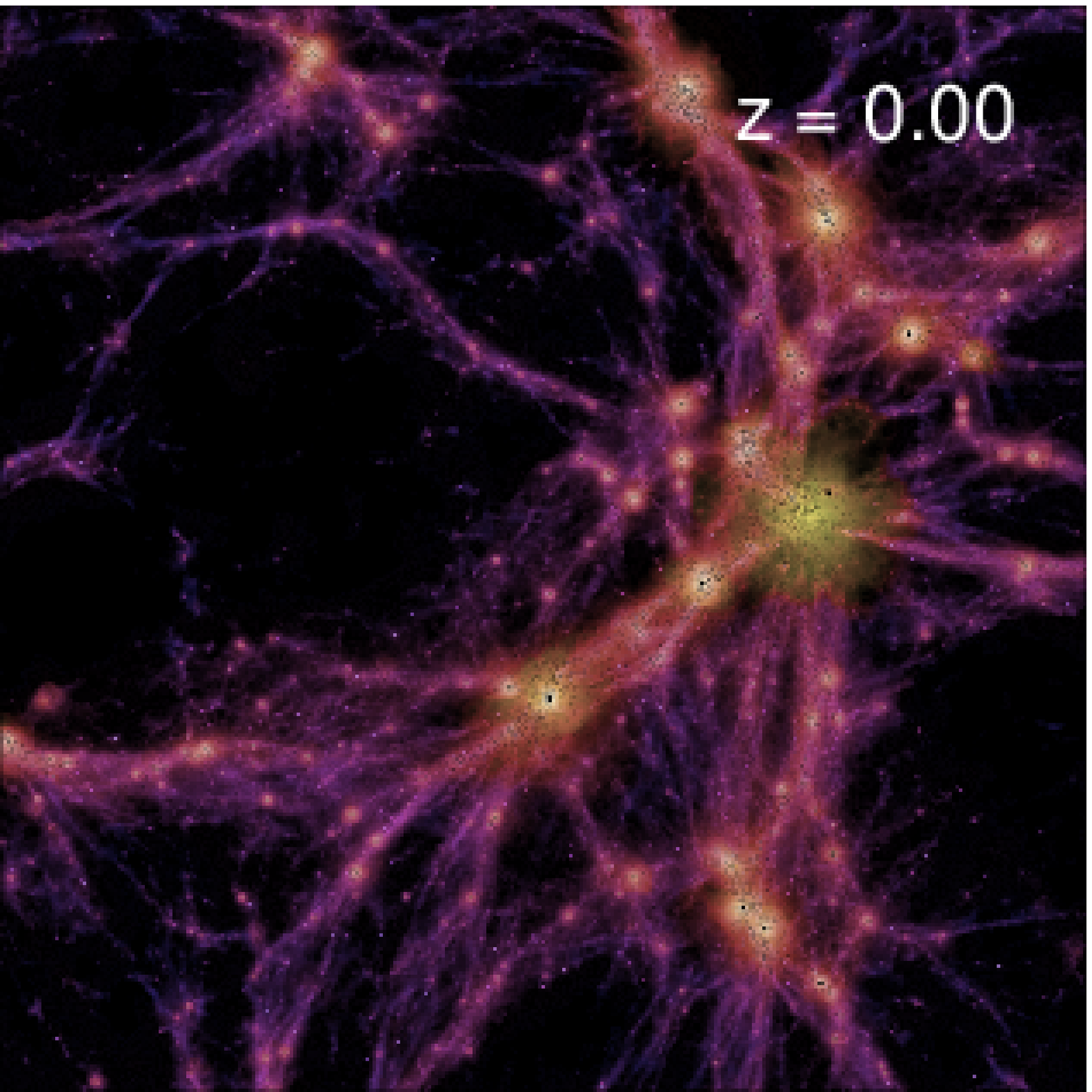}
\caption{A comparison of the 2-D density distribution of baryons (gas, stars and black holes) at various redshifts from three 3-D cosmological simulations with $f_{NL}=0$ (left column), $f_{NL}=100$ (middle column) and $f_{NL}=1000$ (right column), respectively. The region rendered is a spatial slice with a thickness of 10 Mpc/h along Z direction and 50 Mpc/h across in both X and Y directions. For the gas and stars, the brightness corresponds to the density while the color corresponds to the temperature of the gas and the metallicity of the stars. For the color, blue and purple represent the low values (i.e. cold gas and metal poor) while green and yellow represent the high values (i.e. hot gas and metal rich). The black holes are represented in black dots with the size proportional to the black hole mass.}
\label{fg_snap}
\end{center}
\end{figure*}

As a first illustration of the effects of primordial NG on structure formation, Fig. \ref{fg_snap} compares the density distribution of baryons (gas, stars and black holes) at four different redshifts from three simulations with $f_{NL}=0, 100, 1000$, respectively. The NG simulations appear to have different density structures from the Gaussian one, in particular at redshifts $z \gtrsim 2$.  Since a non-Gaussian potential with a positive $f_{NL}$ enhances the density fluctuation, it would affect the collapse time of the overdense regions, the mass of collapsed objects, and the large-scale structure. As shown in Fig. \ref{fg_snap}, the larger the $f_{NL}$ is, the higher the density contrast results, the earlier the structures form, and more objects collapse along the filaments, a trend also found in \cite{Dalal:2008}. 

At hight redshifts ($z \gtrsim 6$), we can clearly distinguish the Gaussian and NG cases even for $f_{NL}=100$, as the latter produce more galaxies early on, and these galaxies appear to cluster strongly. At lower redshift ($z \sim 2$), however, the difference among the three simulations becomes subtle and we can only distinguish the $f_{NL}=1000$ case from the other two through its denser filaments and stronger clustering. By $z =0$, all three cases are almost indistinguishable. Fig. \ref{fg_snap} suggests that the effects of the primordial NG on the large-scale structure tend to be washed away by the non-linear formation and evolution of structures, and the complex feedback processes especially at low redshift. This is consistent with the finding of \cite{Maio:2011} that the NG effects are more significant at higher redshift.  

In what follows, we will quantitatively investigate the effects of primordial NG on galaxy properties, including the abundance, the formation history of stars and black holes, and the clustering and bias. 

\subsection{Galaxy Mass Function}
\label{ss31}

Galaxy abundance is an important aspect used to study the primordial NG as it may affect the formation time and mass of the collapsed objects. In order to determine the NG effects, we compare galaxy mass functions and their evolution produced from different initial conditions, and from simulations with or without baryons to further separate the effects from baryonic processes. 

\begin{figure*}
\begin{center}
\includegraphics[trim=0cm 0cm 0cm 0cm, clip=true, angle=0, width=2.3in]{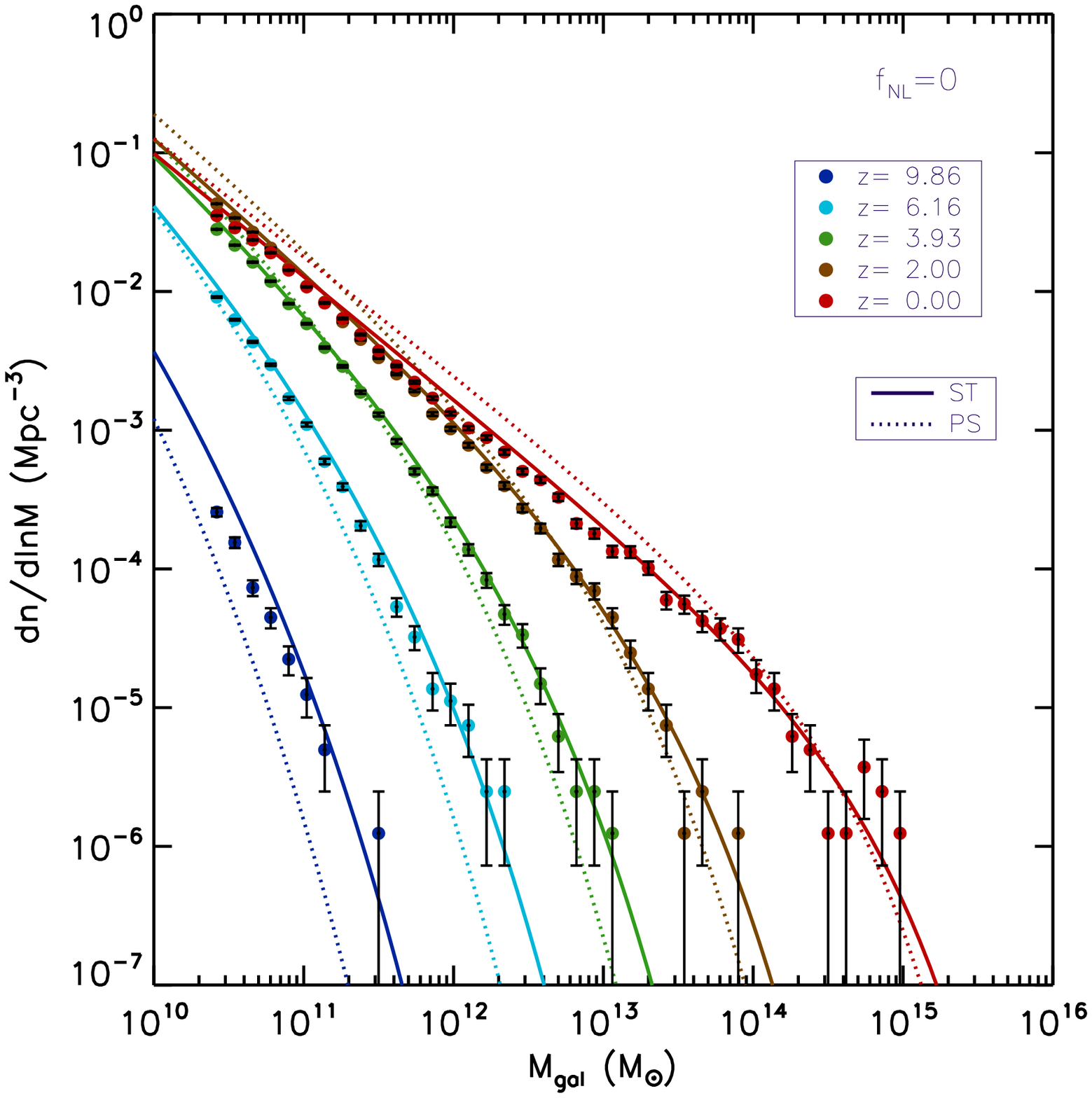}
\includegraphics[trim=0cm 0cm 0cm 0cm, clip=true, angle=0, width=2.3in]{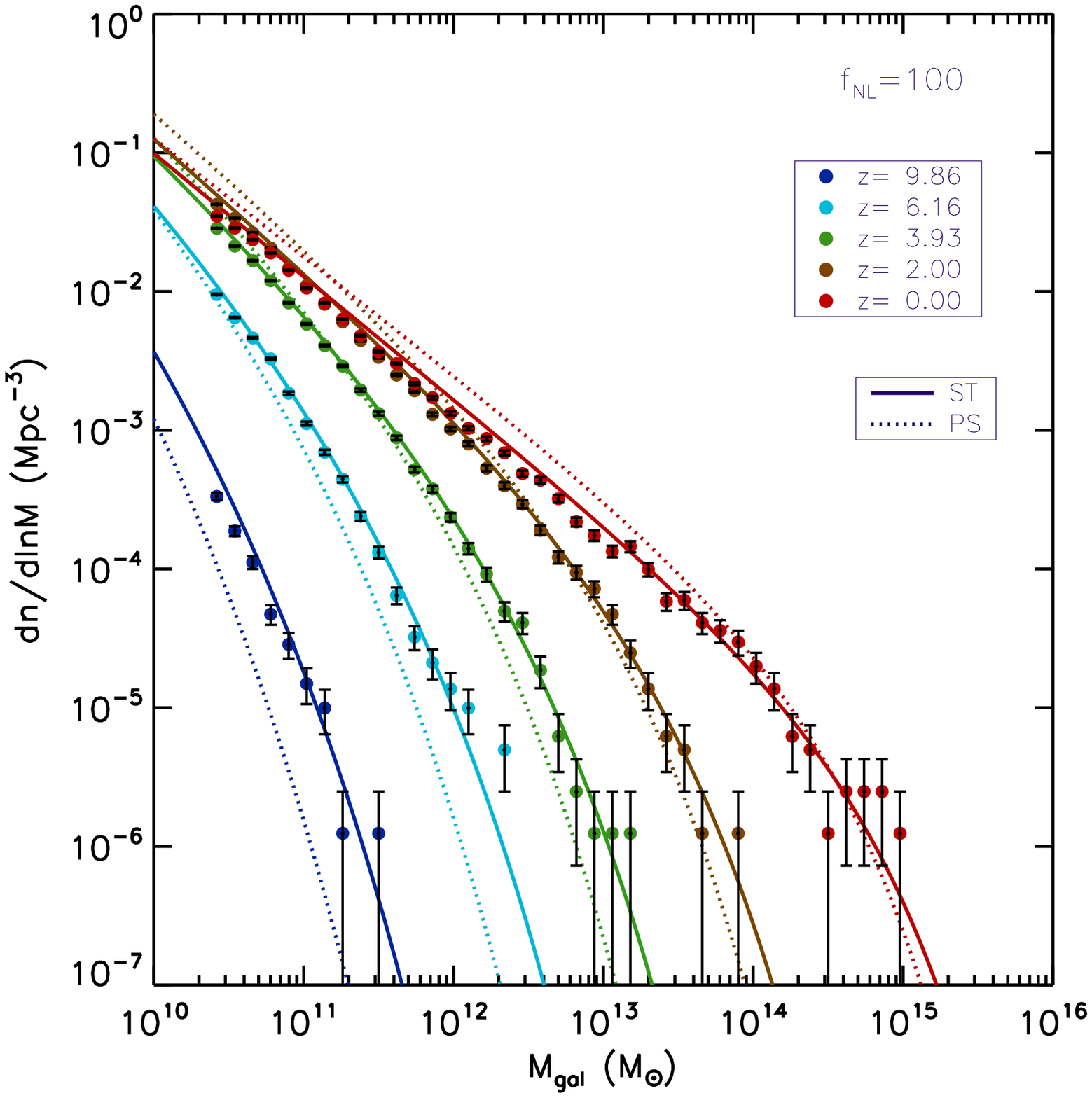}
\includegraphics[trim=0cm 0cm 0cm 0cm, clip=true, angle=0, width=2.3in]{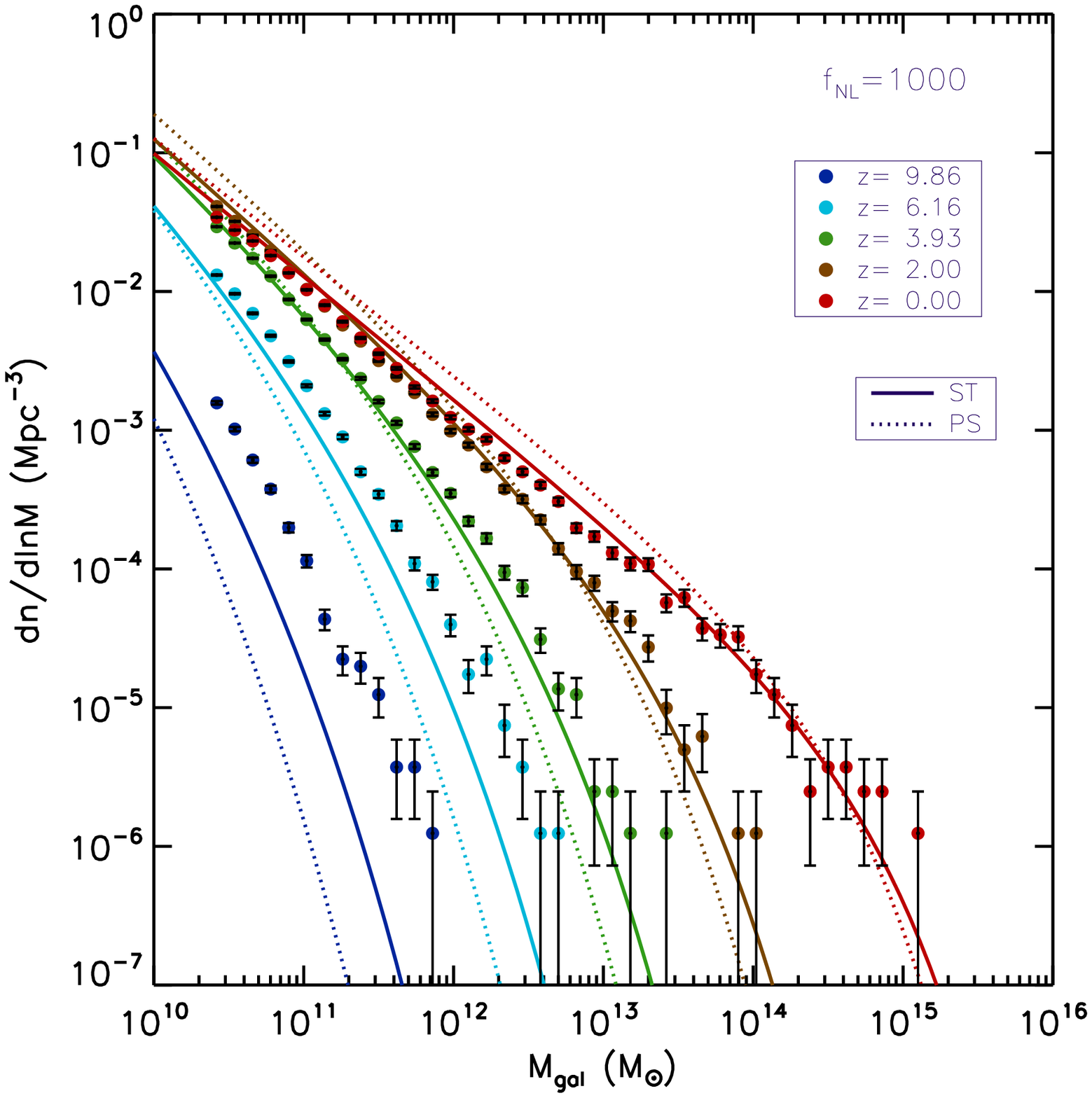} \\
\includegraphics[trim=0cm 0cm 0cm 0cm, clip=true, angle=0, width=2.3in]{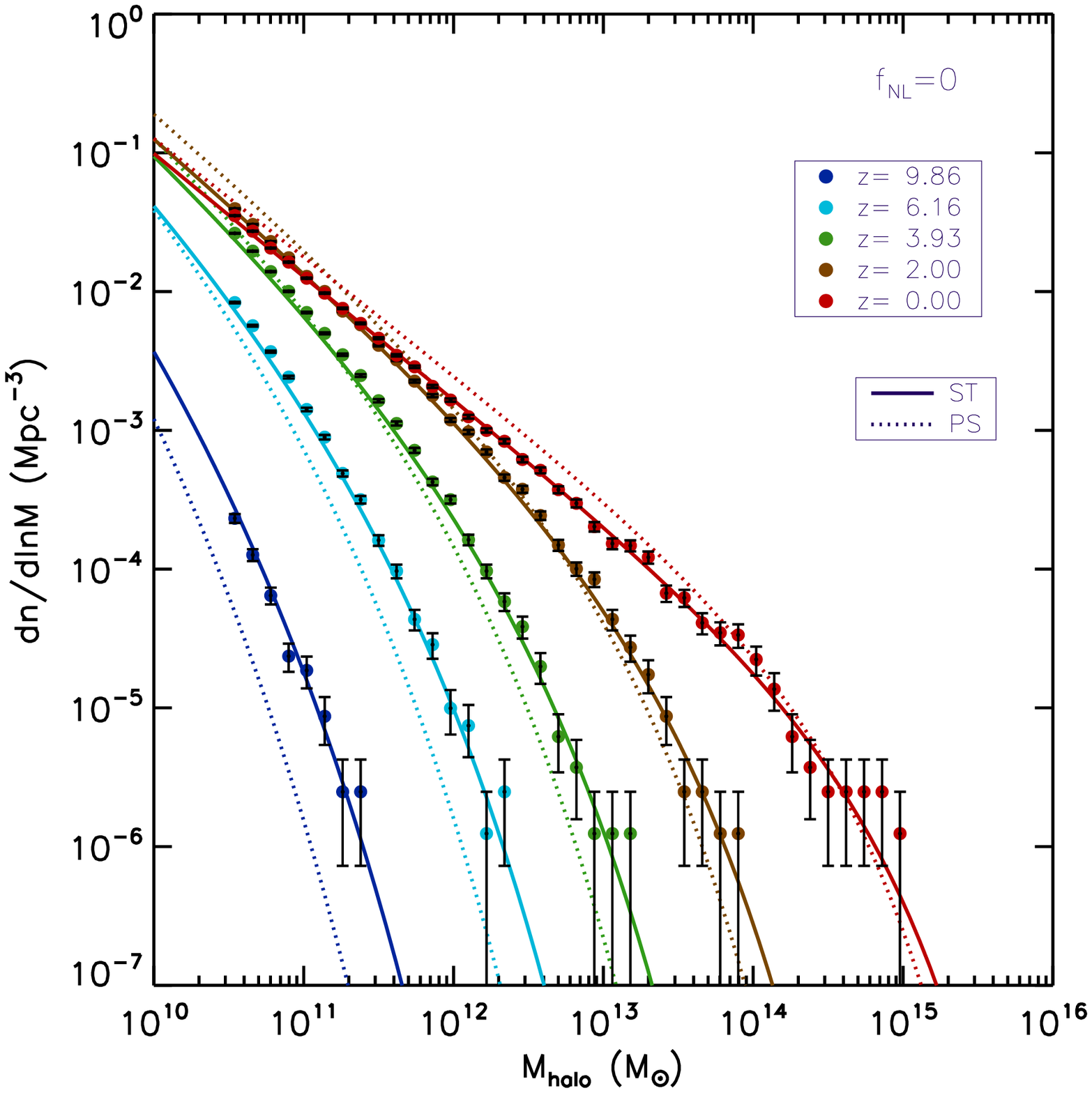}
\includegraphics[trim=0cm 0cm 0cm 0cm, clip=true, angle=0, width=2.3in]{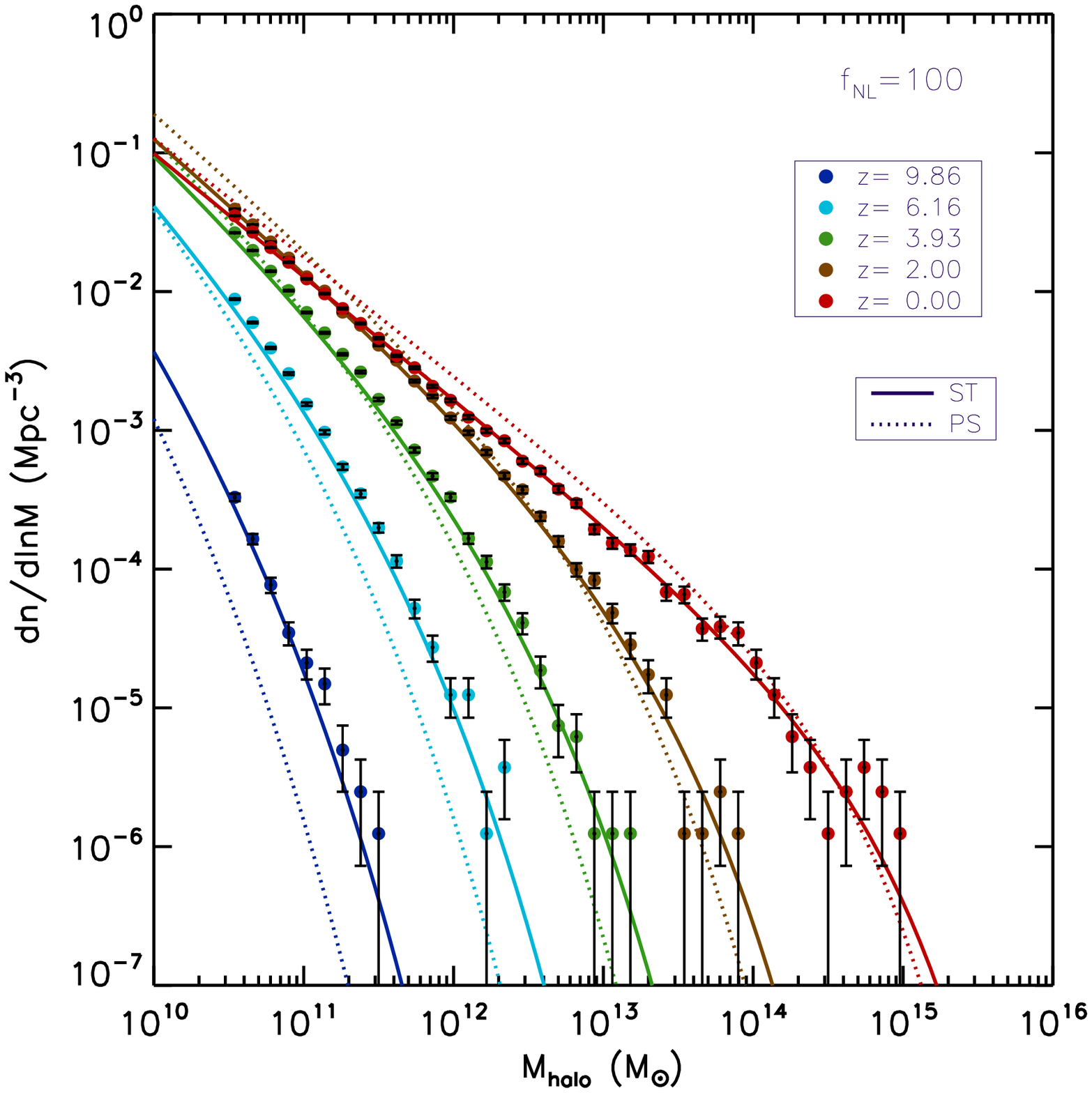}
\includegraphics[trim=0cm 0cm 0cm 0cm, clip=true, angle=0, width=2.3in]{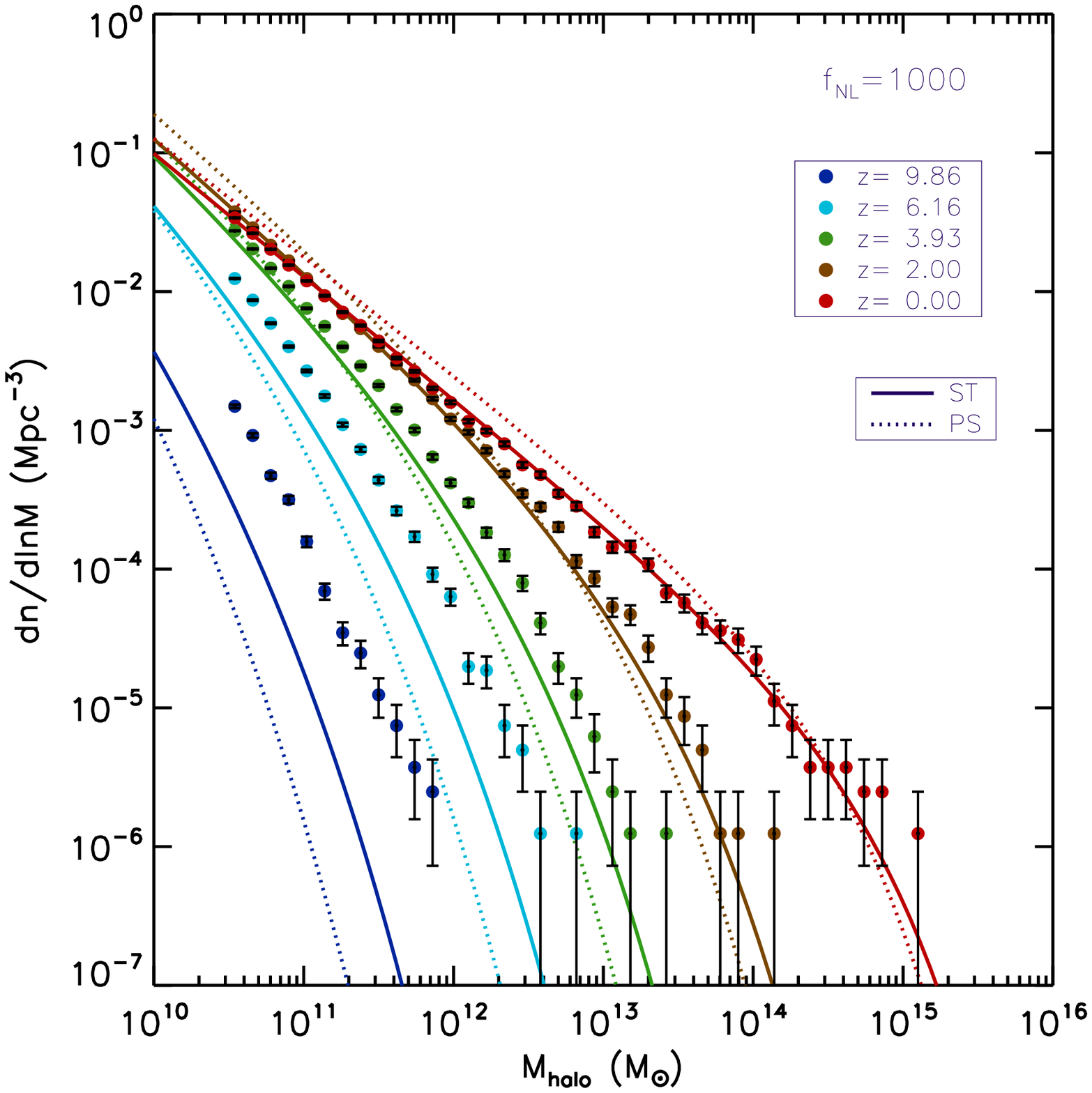}
\caption{A comparison of galaxy mass function at various redshifts from different simulations with $f_{NL}=0, 100, 1000$, respectively. The top row shows the simulations with baryons, while the bottom row shows those with dark matter only. The error bar of each mass bin is the Poission error. Galaxy (Halo) mass function in the baryonic (top row) and dark matter only (bottom row) simulations with $f_{NL}=0, 100, 1000$ at various redshifts along with the theoretical curves as from \cite{Press:1974} (PS) and \cite{Sheth:2002} (ST). }
\label{fg_gal_mass_func}
\end{center}
\end{figure*}

Fig. \ref{fg_gal_mass_func} compares the mass function of galaxies at different redshift from both baryonic and dark matter-only simulations with $f_{NL}=0, 100, 1000$, respectively. As discussed in Sec. \ref{ss22}, we use a two-step group finding procedure to find structures formed in the simulations, so the galaxies (or halos) can be identified by either FOF or SUBFIND algorithm. Since there are still some ambiguities on the criteria to choose one over the other algorithm \citep{Desjacques:2010}, we compare the results returned by both algorithms. Since we find no significant differences between the two in the mass function, and to keep the consistency with some of the previous studies based on dark matter only simulations, we define the galaxies (halos) as the groups identified by the FOF algorithm in the simulations. Since the FOF groups in our simulations have at least 32 dark matter particles, the galaxies used in our data analysis have a minimum mass of $1.79 \times 10^{10}\Msun$/h for dark matter halos. The galaxy mass in Fig. \ref{fg_gal_mass_func} is the total mass of the dark matter, gas, stars and black holes in the group for baryonic simulations, or the sum of all dark matter particles in the group for N-body only simulations

To study the differences in the mass function between the Gaussian and the non-Gaussian cases, we also plot the Gaussian mass functions from the theoretical predictions in \cite{Press:1974} (PS) and \cite{Sheth:2002} (ST) in both Gaussian and non-Gaussian cases in Fig. \ref{fg_gal_mass_func}. As we can see from Fig. \ref{fg_gal_mass_func}, for both baryonic and dark matter only simulations, the mass functions from the simulations agree with the ST prediction reasonably well in the Gaussian case in the redshifts studied. For the non-Gaussian, $f_{NL}=100$, case, it shows a slight excess of the number density of the galaxies at high redshifts ($z \ge 6$). But for the lower redshifts, the difference between the Gaussian and non-Gaussian cases is very subtle and they are almost indistinguishable. For the non-Gaussian, $f_{NL}=1000$, case, the simulations clearly show a substantial derivation from the Gaussian case in the mass function. First, it shows a large excess of the number density of the galaxies at all redshifts except $z=0$ and the excess increases with redshift. And second, more massive galaxies can form in this case than in either the Gaussian or $f_{NL}=100$ case. We note, however, that the most massive galaxy at $z=0$ in the $f_{NL}=1000$ case is only slightly more massive than that in the Gaussian and $f_{NL}=100$ cases. The main difference between the baryonic and dark matter only simulations is that the number density of the halos is slightly higher than that of the galaxies in all cases so that it agrees with the ST prediction better as expected (both PS and ST formalisms are based on the dark matter halo model). Fig. \ref{fg_gal_mass_func} shows that the effects of the NG on the galaxy population are substantial at high redshifts and increase with the magnitude of the NG. These effects are more subtle at low redshifts even for the very non-Gaussian ($f_{NL}=1000$) case.

The theoretical formulation of the mass function from non-Gaussian initial conditions has been presented by many studies (\cite{Matarrese:2000,Dalal:2008,LoVerde:2008,Pillepich:2008ka,Maggiore:2010,LoVerde:2011}). We compare our simulation results to the non-Gaussian mass functions from \cite{Matarrese:2000} and \cite{LoVerde:2008}. Both of the two formulations use extensions of the Press-Schechter formalism with slightly different approximations made in the derivations. In the limit of very small NG, the results from both derivations agree with each other. However, both formulations are limited by the several assumptions made in their derivations (spherical collapse, sharp k-space filtering and random-phase hypothesis) and they are only good for mild NG cases. Because of these issues in the two derivations, it is argued \citep{Verde:2001,LoVerde:2008,Grossi:2009} that one should calculate the non-Gaussian mass function by using the Gaussian mass function that is well studied and calibrated by numerical simulations multiplied by a relative non-Gaussian correction factor. We will use this approach to further study the non-Gaussian mass function by comparing the galaxy (halo) mass functions from the simulations with the predictions from \cite{Matarrese:2000} and \cite{LoVerde:2008}.

\begin{figure*}
\begin{center}
\includegraphics[trim=0cm 0cm 0cm 0cm, clip=true, angle=0, width=2.3in]{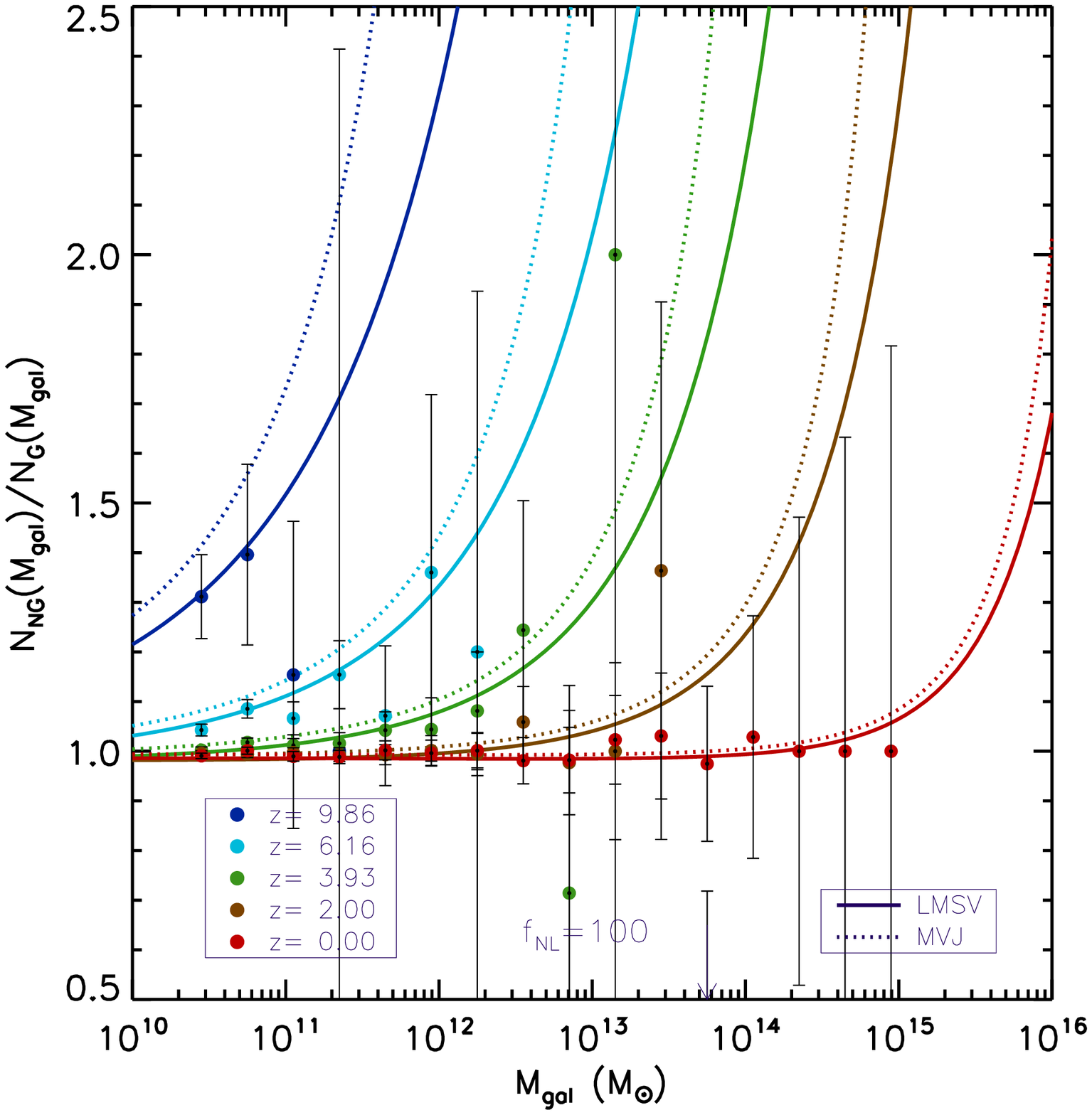}
\includegraphics[trim=0cm 0cm 0cm 0cm, clip=true, angle=0, width=2.3in]{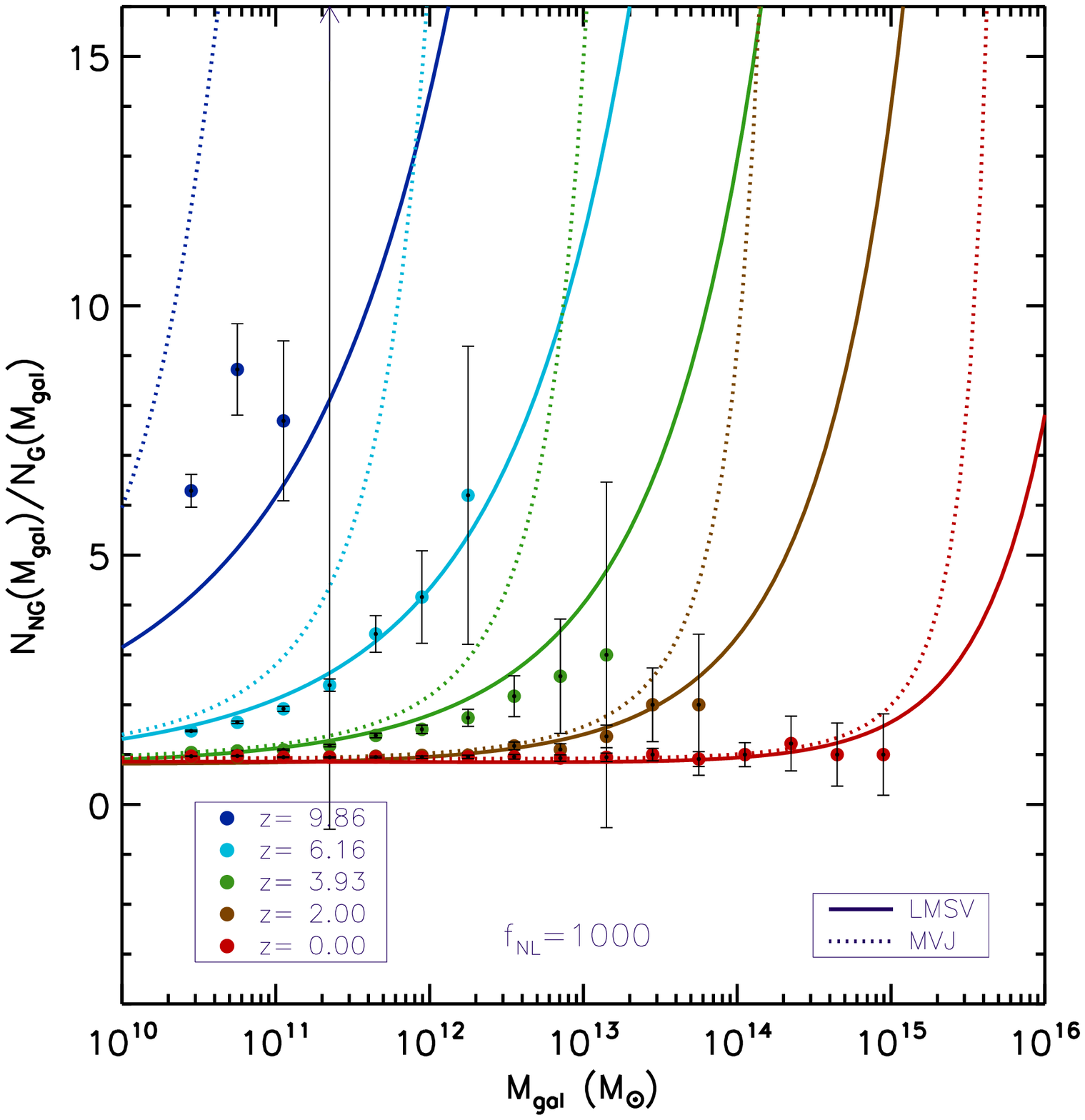} \\
\includegraphics[trim=0cm 0cm 0cm 0cm, clip=true, angle=0, width=2.3in]{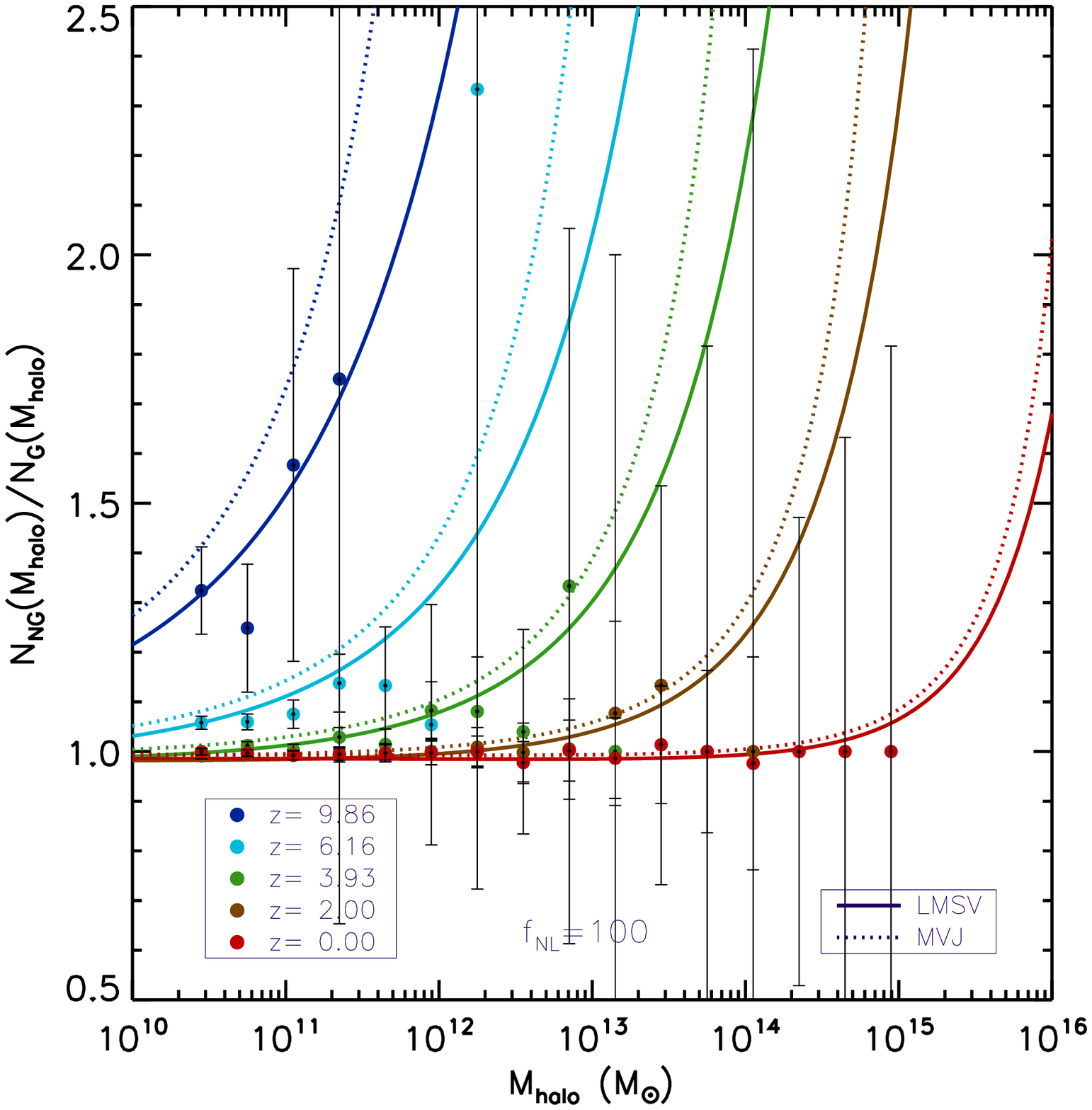}
\includegraphics[trim=0cm 0cm 0cm 0cm, clip=true, angle=0, width=2.3in]{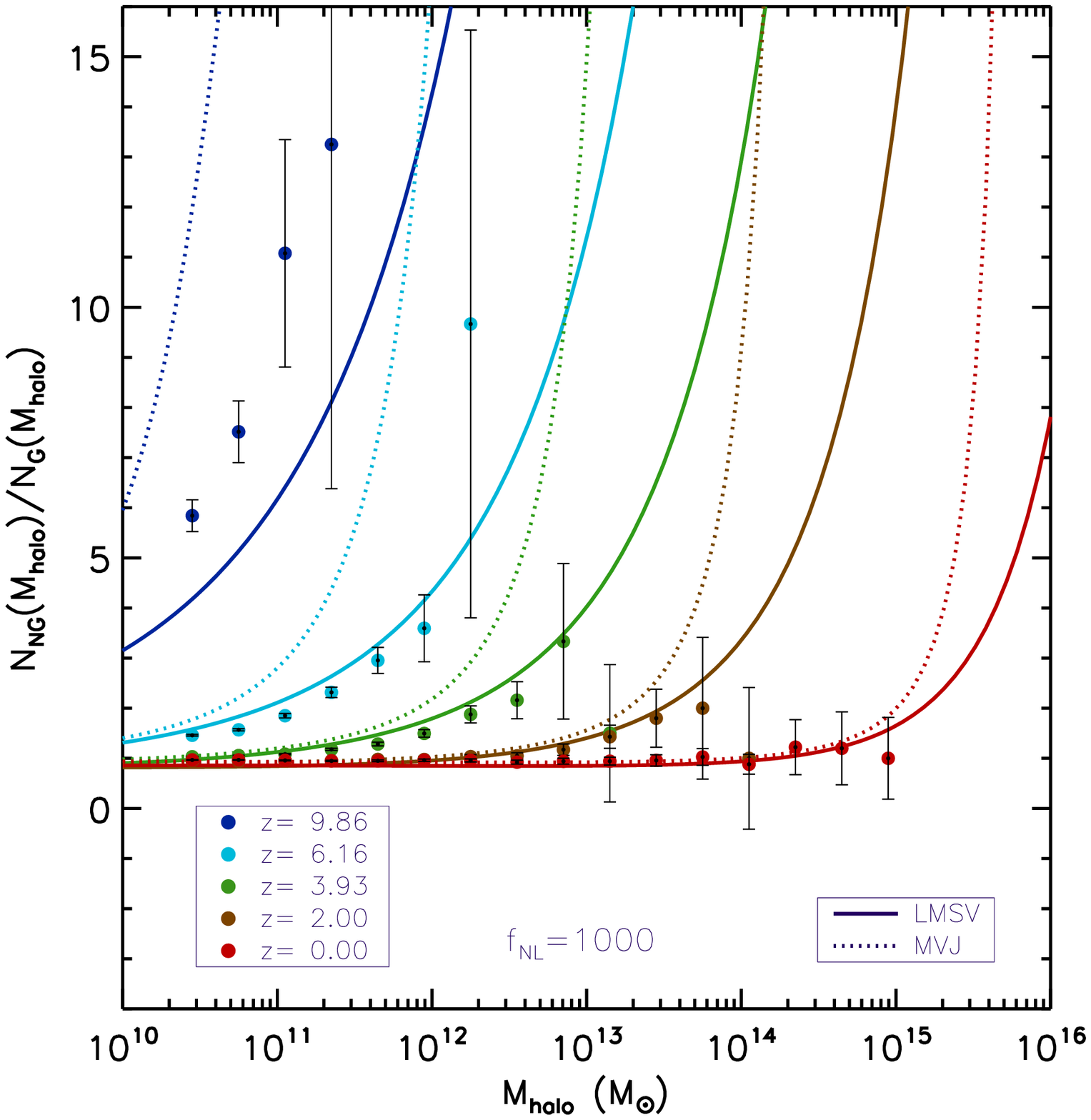}
\caption{The ratio between the galaxy (halo) mass functions of $f_{NL}=100$ and $f_{NL}=0$ cases (left) and $f_{NL}=1000$ and $f_{NL}=0$ cases (right) at various redshifts for the baryonic (top row) and dark matter only (bottom row) simulations along with the theoretical predictions from \cite{Matarrese:2000} (MVJ) and \cite{LoVerde:2008} (LMSV).}
\label{fg_gal_mass_func_ratio}
\end{center}
\end{figure*}

In Fig. \ref{fg_gal_mass_func_ratio}, we plot the ratio of the galaxy (halo) mass functions between non-Gaussian ($f_{NL}=100, 1000$) and Gaussian ($f_{NL}=0$) cases for the baryonic simulations in the top row and dark matter only simulations in the bottom row at various redshifts. 
We write the non-Gaussian mass function $n_{NG}(M,z,f_{NL})$ as:
\begin{equation}
n_{NG}(M,z,f_{NL})=n_G(M,z,f_{NL}=0)R(M,z,f_{NL}),
\end{equation}
where $n_G(M,z,f_{NL}=0)$ is the Gaussian mass function that can be obtained from numerical simulations and $R(M,z,f_{NL})$ is the non-Gaussian correction factor. $M$ corresponds to the smoothing scale $R$ of the density field so that $M=4/3\pi R^3 \rho$. The exact forms of $R(M,z,f_{NL})$ can be extracted from the expressions of the non-Gaussian mass functions in \cite{Matarrese:2000} and \cite{LoVerde:2008}. So $R(M,z,f_{NL})$ can be written either as from \cite{Matarrese:2000} (MVJ):
\begin{eqnarray}
&&R(M,z,f_{NL})={\rm{exp}} \left[ \delta_{ec}^3\frac{S_{3,M}}{6\sigma_M^2} \right] \times \nonumber \\
&&\left| \frac{1}{6}\frac{\delta_{ec}}{\sqrt{1-\delta_{ec}S_{3,M}/3}} \frac{{\rm{d}} S_{3,M}}{{\rm{d ln}} \sigma_M} + \sqrt{1-\delta_{ec}S_{3,M}/3} \right|,
\end{eqnarray}
or as from \cite{LoVerde:2008} (LMSV):
\begin{eqnarray}
&&R(M,z,f_{NL})=1+ \frac{1}{6} \frac{\sigma_M^2}{\delta_{ec}} \times \nonumber \\
&&\left[ S_{3,M}\left(\frac{\delta_{ec}^4}{\sigma_M^4}-2\frac{\delta_{ec}^2}{\sigma_M^2}-1\right) + \frac{{\rm{d}} S_{3,M}}{{\rm{d ln}} \sigma_M}\left( \frac{\delta_{ec}^2}{\sigma_M^2}-1 \right) \right],
\end{eqnarray}
where $\delta_{ec}$ is the critical overdensity of ellipsoidal collapses. For the high peaks limit, $\delta_{ec}=\delta_c\sqrt{q}$, where $\delta_c$ is the critical overdensity for spherical collapses and $q \le 1$ is a numerical factor that is normally calibrated by numerical simulations. The exact value of $q$ depends on the halo finding algorithm, such as FOF or spherical overdensity (SO) \citep{Desjacques:2010,Wagner:2010}, and the theoretical formulation \citep{Matarrese:2000,LoVerde:2008,LoVerde:2011}. Since we define the galaxies (halos) as the particle groups identified by the FOF algorithm with the linking length of 0.2 times of the mean particle spacing in the simulations, we take $q$ to be 0.75 in this study. $S_{3,M}$ is the normalized skewness of the density field with a smoothing scale corresponding to mass M and $S_{3,M}=\langle\delta^3_M\rangle/\langle\delta^2_M\rangle^2$, where $\langle\delta^3_M\rangle$ and $\langle\delta^2_M\rangle$ are the 3rd and 2nd moment of the density field at this scale. $\sigma_M^2=\langle\delta^2_M\rangle$ is the variance of the smoothed density field. We calculate $S_{3,M}$ with a fitting formula from \cite{Figueroa:2012} as:
\begin{equation}
S_{3,M}=\frac{f_{NL}}{\sigma_M}{\rm{exp}}(-7.98-0.177{\rm{ln}}\nu+0.0186{\rm{ln}}^2\nu-0.00260{\rm{ln}}^3\nu),
\end{equation}
where $\nu=\delta_{ec}/\sigma_M$. We overplot the mass functions as derived in MVJ and LMSV in Fig. \ref{fg_gal_mass_func_ratio} to illustrate how well they agree with the mass functions extracted from our simulations.

As we can see from the first row of Fig. \ref{fg_gal_mass_func_ratio}, the mass function ratio $N_{NG}(M_{gal})/N_{G}(M_{gal})$ for the baryonic simulations agrees reasonably well with the theoretical predictions from MVJ and LMSV as found that of dark matter only simulations in the previous studies \citep{Matarrese:2000,LoVerde:2008,Grossi:2009}. For the $f_{NL}=100$ case, the simulation and both theoretical predictions agree best at intermediate redshifts ($z=6.16$ and $z=3.93$) and still reasonably well at high redshift ($z=9.86$) and low redshifts ($z=2$ and $z=0$) although the simulation data shows some large scatter ($z=9.86$ and $z=2$) and the difference between the Gaussian and non-Gaussian cases is subtle at the latest time ($z=0$). For the $f_{NL}=1000$ case, the LMSV prediction shows a better agreement with the simulation data at low and intermediate redshifts in the low to medium mass range. Both theoretical predictions show substantial deviations from the simulation data at high redshift and the high mass end. This may be expected in the case where the LMSV prediction is truncated at first order (skewness only) which is not accurate for large $f_{NL}$, high redshifts and very massive objects. But surprisingly, the MVJ prediction that is supposed to be a good approximation for rare massive objects \citep{Verde:2010} doesn't perform substantially better in these cases either. Although it has been found \citep{Dalal:2008} that the MVJ prediction over predicts the number density of the halos in the non-Gaussian simulations as we find in this study, we think more studies are needed to confirm this aspect in baryonic simulations. Our results are somewhat limited by the small box size and small number of realizations.

For the halo mass function ratio $N_{NG}(M_{halo})/N_{G}(M_{halo})$ of the dark matter only simulations in the bottom row of Fig. \ref{fg_gal_mass_func_ratio}, it shows a slightly better agreement with the theoretical predictions from MVJ and LMSV. But the improvement over the baryonic simulations is just marginal. So for the galaxy mass function in the simulations with baryonic matter, the MVJ and LMSV formulations can give comparable approximate predictions of its overall trend in the data range investigated in this study like they do for the dark matter only simulations. The additional baryonic physics doesn't have a substantial impact on the mass function of the galaxies. We will study the stellar component next.

\subsection{Star Formation History}
\label{ss32}

\begin{figure}
\begin{center}
\includegraphics[angle=0, width=3.4in]{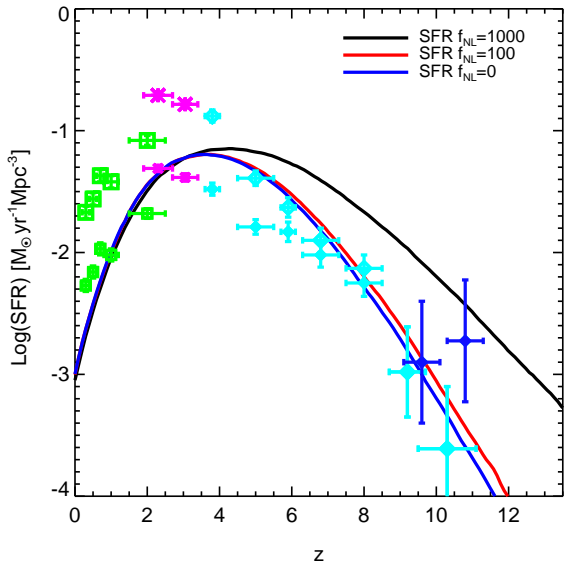}
\caption{The cosmic star formation history derived from the simulations of $f_{NL}=0, 100, 1000$, respectively, in comparison with observations. The observations include data from  \cite{Schiminovich:2005} (green points), \cite{Reddy:2009} (magenta points), \cite{Bouwens:2007,Bouwens:2009,Bouwens:2012} (cyan points) and \cite{Coe:2013} (blue points). For double data points of one redshift, the upper, bigger symbol corresponds to the dust-corrected data as presented in \cite{Bouwens:2012}.}
\label{fg_sfr_bhar}
\end{center}
\end{figure}

The impact of the primordial NG on the cosmic star formation history is illustrated in Fig. \ref{fg_sfr_bhar}, where we plot the cosmic star formation rate density  from the Gaussian ($f_{NL}=0$) and the non-Gaussian ($f_{NL}=100, 1000$) simulations. We can clearly see that for a simulation with a positive $f_{NL}$, the stars start to form both earlier and faster at high redshifts (beyond $z\sim4$). The star formation rate density in both Gaussian and non-Gaussian simulations peaks between redshift $3<z<4$. And for low redshifts $z\le3$, it is almost impossible to distinguish between Gaussian and non-Gaussian cases. The differences in the star formation history between the non-Gaussian, $f_{NL}=1000$, and Gaussian $f_{NL}=0$ cases are quite substantial at high redshifts. The star formation rate density is almost two orders of magnitude higher at $z\sim20$ in the $f_{NL}=1000$ case than the $f_{NL}=0$ case. The stars also start to form much earlier in the $f_{NL}=1000$ case ($z\sim27$ for $f_{NL}=1000$ versus $z\sim20$ for $f_{NL}=0$). The differences between the $f_{NL}=100$ and $f_{NL}=0$ cases are not substantial except at high redshifts (beyond $z\sim10$). And moreover, the biggest difference in the star formation rate density is much less than one order of magnitude in the full history in these two cases and the stars start to form around the same time too ($z\sim20$). Again the star formation history is nearly identical for all three cases for low redshifts $z\le3$. This is consistent with the trend found in the recent studies by \cite{Maio:2011}, \cite{Maio:2011a} and \cite{Maio:2012a}. So if we want to use the star formation rate from the observations to constrain the primordial NG, we need to use the high redshift data to break the degeneracy.

In order to compare the cosmic star formation rate density from our Gaussian and non-Gaussian simulations with the one obtained from the astronomical observations, in the bottom panel of Fig. \ref{fg_sfr_bhar}, we overplot the observational data from recent studies \citep{Schiminovich:2005,Reddy:2009,Bouwens:2007,Bouwens:2009,Bouwens:2012,Coe:2013} along the data from our simulations. For redshift $z \le 8$, we also include the dust-corrected data in \cite{Bouwens:2012}. As we can see from Fig. \ref{fg_sfr_bhar}, the data from our $f_{NL}=0$ and $f_{NL}=100$ simulations agree with the observations relatively well for redshift $z>4$. For the $f_{NL}=1000$ case, the simulation seems to over produce the stars by a wide margin. The star formation rate density data from all three simulations agree with the original observational data roughly at redshift $z\le4$ although they appear to peak a little earlier ($3<z<4$) than the ones from the observations ($z\sim2.5$). The simulation data are also systematically lower than the dust-corrected data from the observations at redshift $z\le4$. However, this doesn't affect our ability to use the simulation and observational data to constrain the primordial NG because at low redshifts, the star formation rate density from the simulations is very sensitive to the galactic wind strength in our supernova feedback model that is not well constrained from current observations. In addition, the current dust-correction model still has several uncertainties. Finally, and most importantly, the star formation rate density is degenerate for all three non-Gaussian cases at these redshifts, so we are only interested in the star formation rate density at high redshifts to constrain the primordial NG. However, in order to use the data from the simulations to give a tight constraint, there are still many factors need to be considered. For example, the mass resolution may affect the star formation rate at high redshifts as discussed in \cite{Springel:2003a}. This is probably one of the factors that cause the slight difference between the star formation history derived from the simulations in this study and the one presented in \cite{Maio:2011}. So future detailed studies on the star formation model are needed to calibrate the simulation data with the observations on this aspect.

\begin{figure*}
\begin{center}
\includegraphics[trim=0cm 0cm 0cm 0cm, clip=true, angle=0, width=2.3in]{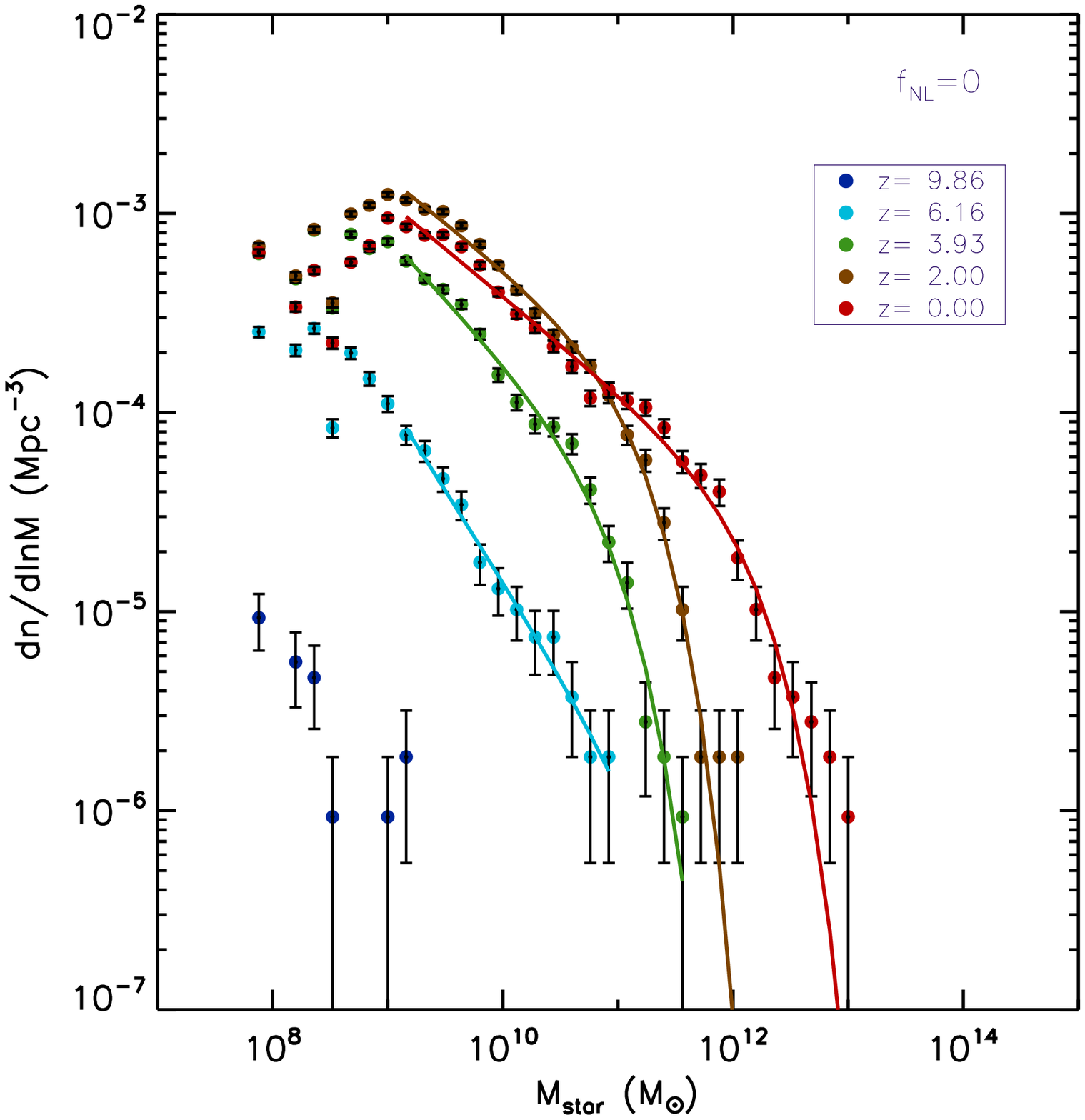}
\includegraphics[trim=0cm 0cm 0cm 0cm, clip=true, angle=0, width=2.3in]{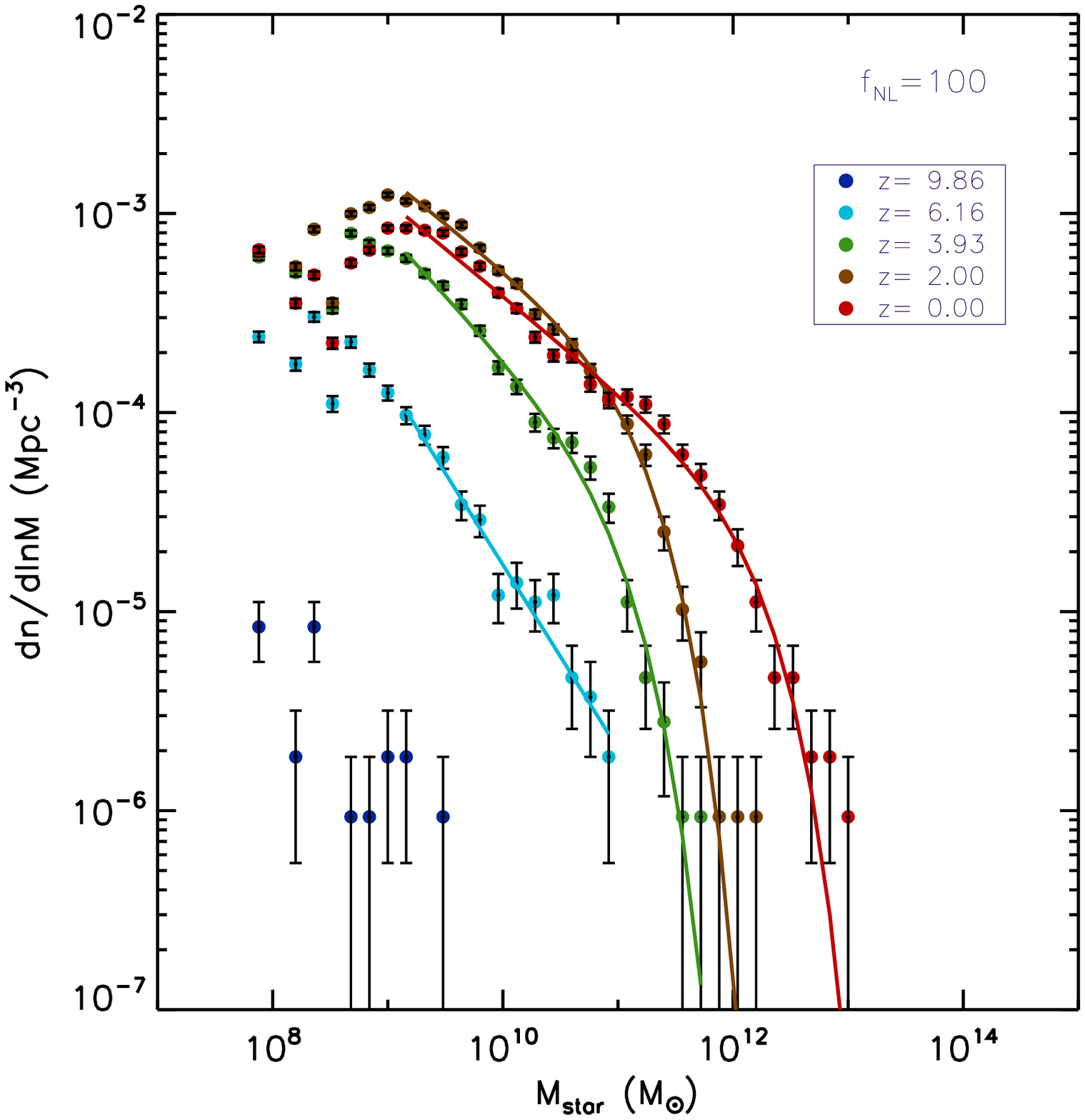}
\includegraphics[trim=0cm 0cm 0cm 0cm, clip=true, angle=0, width=2.3in]{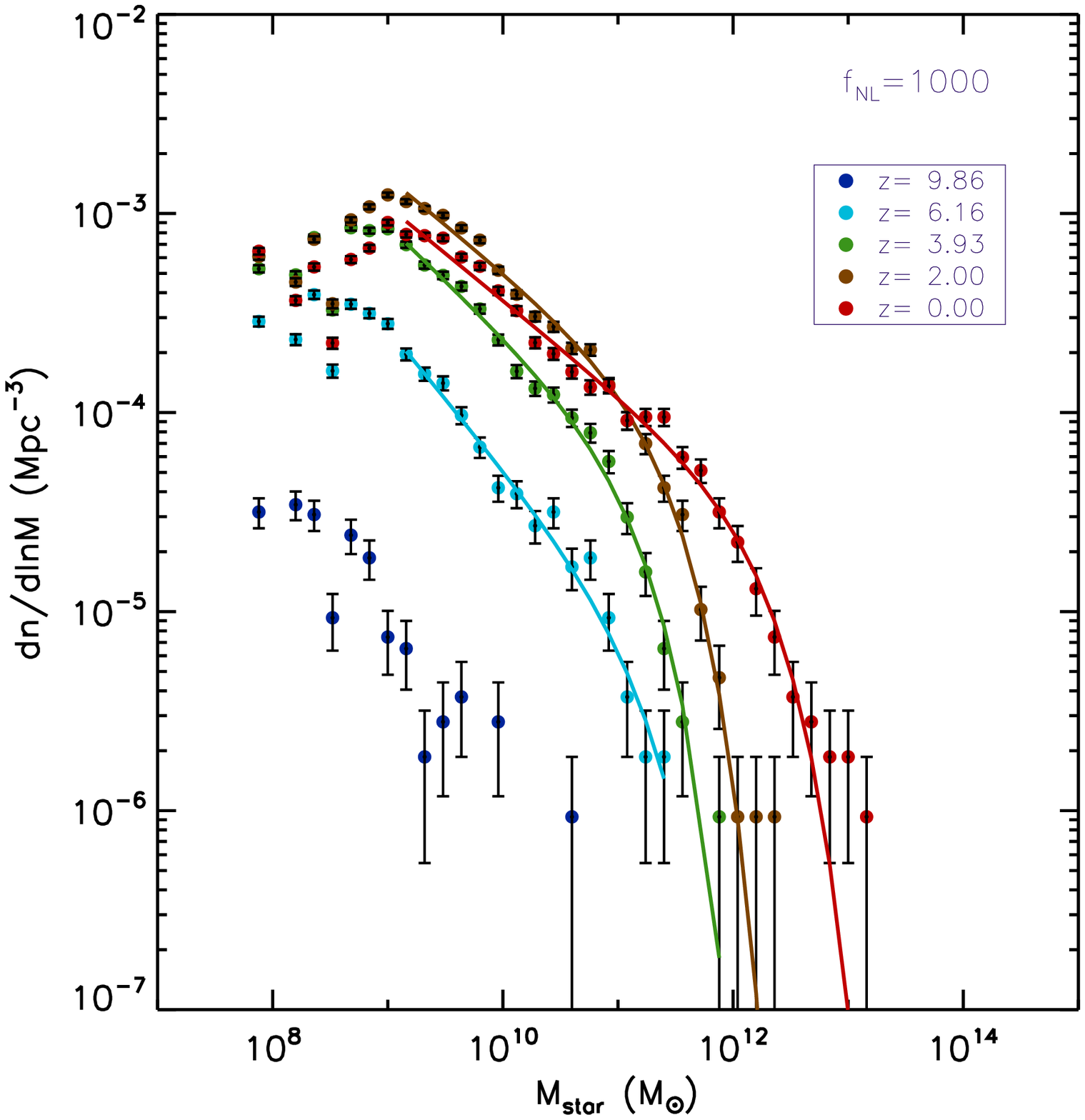}
\caption{The stellar mass function of galaxies at various redshifts from simulations with $f_{NL}=0, 100, 1000$, respectively. A Schechter function is used as the fitting curve to the simulation data at low redshifts ($z <4$), while a power law is used for high-redshift data ($z >4$).}
\label{fg_stellar_mass_func}
\end{center}
\end{figure*}

\begin{figure}
\begin{center}
\includegraphics[angle=0, width=3.4in]{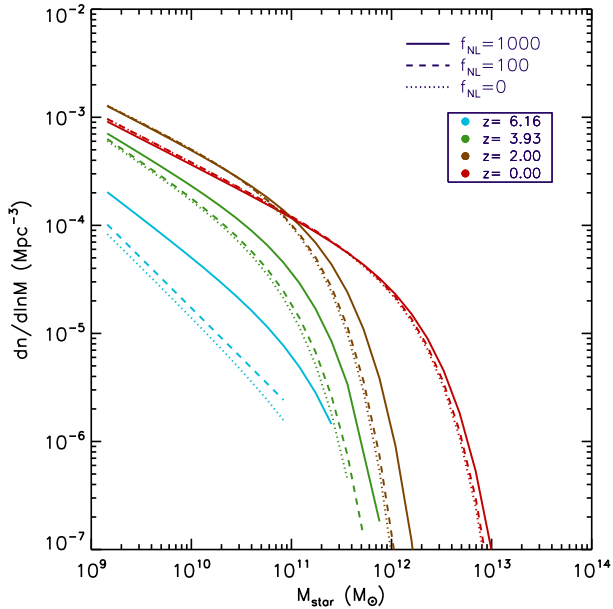}
\caption{A comparison of the fitting curves of the stellar mass functions of galaxies for all $f_{NL}=0, 100, 1000$ simulations as shown in Figure~ \ref{fg_stellar_mass_func}.}
\label{fg_stellar_mass_func_comp}
\end{center}
\end{figure}

The impacts of the primordial NG on the luminous components of galaxies is illustrated in Fig. \ref{fg_stellar_mass_func}, where we plot the stellar mass function of galaxies of the Gaussian ($f_{NL}=0$) and the non-Gaussian ($f_{NL}=100, 1000$) simulations at various redshifts. The stellar mass in this plot corresponds to the total mass of the stellar particles in each galaxy in the simulations and the galaxy population in this plot is same as that in Fig. \ref{fg_gal_mass_func}. In order to study the differences between the Gaussian and non-Gaussian simulations, we also fit the stellar mass function using a Schechter luminosity function \citep{Schechter:1976},
\begin{equation}
\phi(L){\rm{d}}L=\phi^*(\frac{L}{L^*})^\alpha {\rm{exp}}(-\frac{L}{L^*})\frac{{\rm{d}}L}{L^*},
\label{eq_schechter_func}
\end{equation}
where $\phi(L)$ is the number density for a galaxy luminosity $L$, $\phi^*$ is a normalization factor, $L^*$ is a characteristic galaxy luminosity where a power law at the faint end transitions to an exponential drop at the bright end and $\alpha$ is a faint end slope. We fit the stellar mass of galaxies as their luminosity in the Schechter function and overplot the fitted curves in the same figure. To reduce the numerical error when galaxies only contain very few stellar particles, we only include the galaxies with at least 20 stellar particles in the data fitted. There are too few galaxies satisfying this criterion at $z=9.86$, so we don't include this redshift for the fitting. For an easy comparison between the Gaussian and non-Gaussian cases, in Fig. \ref{fg_stellar_mass_func_comp}, we plot the fitted curves in all $f_{NL}=0, 100, 1000$ simulations in one figure.

As we can see from the stellar mass function in Figure~\ref{fg_stellar_mass_func} and Fig. \ref{fg_stellar_mass_func_comp}, for the non-Gaussian, $f_{NL}=100$, case, it shows a slightly higher number density than the Gaussian case at all redshifts plotted. The excess of the number density is bigger at high redshifts and for more luminous galaxies. At low redshifts and for less luminous galaxies, the excess becomes smaller and for the low luminous ends at $z=2.00$ and $z=0.00$, the Gaussian and non-Gaussian case are almost indistinguishable. For the $f_{NL}=1000$ case, it shows a similar trend as the $f_{NL}=100$ case. But it has a much bigger excess of the number density than the $f_{NL}=100$ case especially at high redshifts. The excess increases with redshift and again at the low luminousity ends at low redshifts, the difference between the Gaussian and non-Gaussian case is very small. We think the properties of the non-Gaussian stellar mass function as we illustrate here are at least partially due to the combination of the non-Gaussian effects on the galaxy mass function (as shown in Fig. \ref{fg_gal_mass_func} and Fig. \ref{fg_gal_mass_func_ratio}) and star formation rate (as shown in Fig. \ref{fg_sfr_bhar}). To quantitatively calculate the non-Gaussian stellar mass function and to further compare it with the observational data, a more detailed study with additional radiative transfer calculations is needed. We will include it in our future investigations.

\subsection{Black Hole Growth History}
\label{ss33}

\begin{figure}
\begin{center}
\includegraphics[angle=0, width=3.4in]{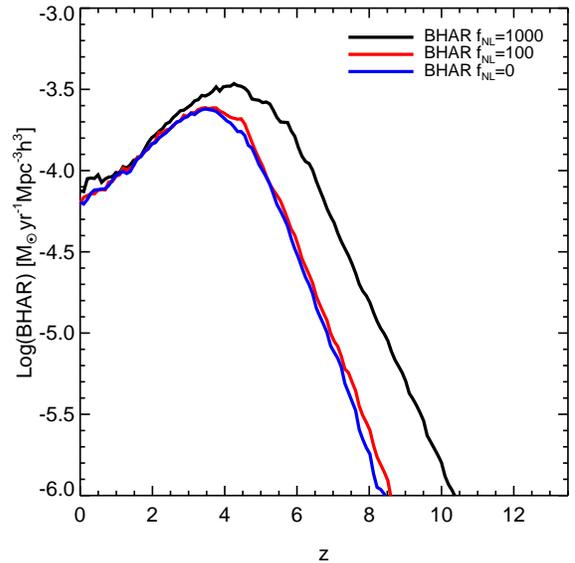} 
\caption{The evolution of black hole accretion rate density derived from the $f_{NL}=0, 100, 1000$ simulations, respectively. }
\label{fg_bhar}
\end{center}
\end{figure}

Similar to the study of the stellar component presented in Sec. \ref{ss32}, we now study the impact of the primordial NG on the black hole accretion (growth) history. As shown in Fig. \ref{fg_bhar}, the black holes start to form earlier and accrete faster at high redshifts ($z>4$) in a simulation with a positive $f_{NL}$. The black hole accretion rate density in both Gaussian and non-Gaussian simulations peaks at redshift $3<z\le4$ and for low redshifts $z\le3$, it is almost impossible to distinguish between Gaussian and non-Gaussian cases. The difference in the black hole accretion history between the $f_{NL}=1000$ and $f_{NL}=0$ cases is quite substantial at high redshifts ($z>4$). The black hole accretion rate density in the $f_{NL}=1000$ case is more than two orders of magnitude higher at $z\sim14$ than that in the $f_{NL}=0$ case although the black holes start to form just a little earlier in the $f_{NL}=1000$ case ($z\sim17$) than that in the $f_{NL}=0$ case ($z\sim14$). The difference between the $f_{NL}=100$ and $f_{NL}=0$ cases is very small in the full redshift range studied and the black holes start to form around the same time ($z\sim14$) in both cases. Just like the stellar component, the black hole accretion history is almost same for all three cases for low redshifts $z\le3$. So, to use the black hole accretion rate to constrain the primordial NG, we need to use astronomical observables related to the black holes at high redshifts such as quasars.

\begin{figure*}
\begin{center}
\includegraphics[trim=0cm 0cm 0cm 0cm, clip=true, angle=0, width=2.3in]{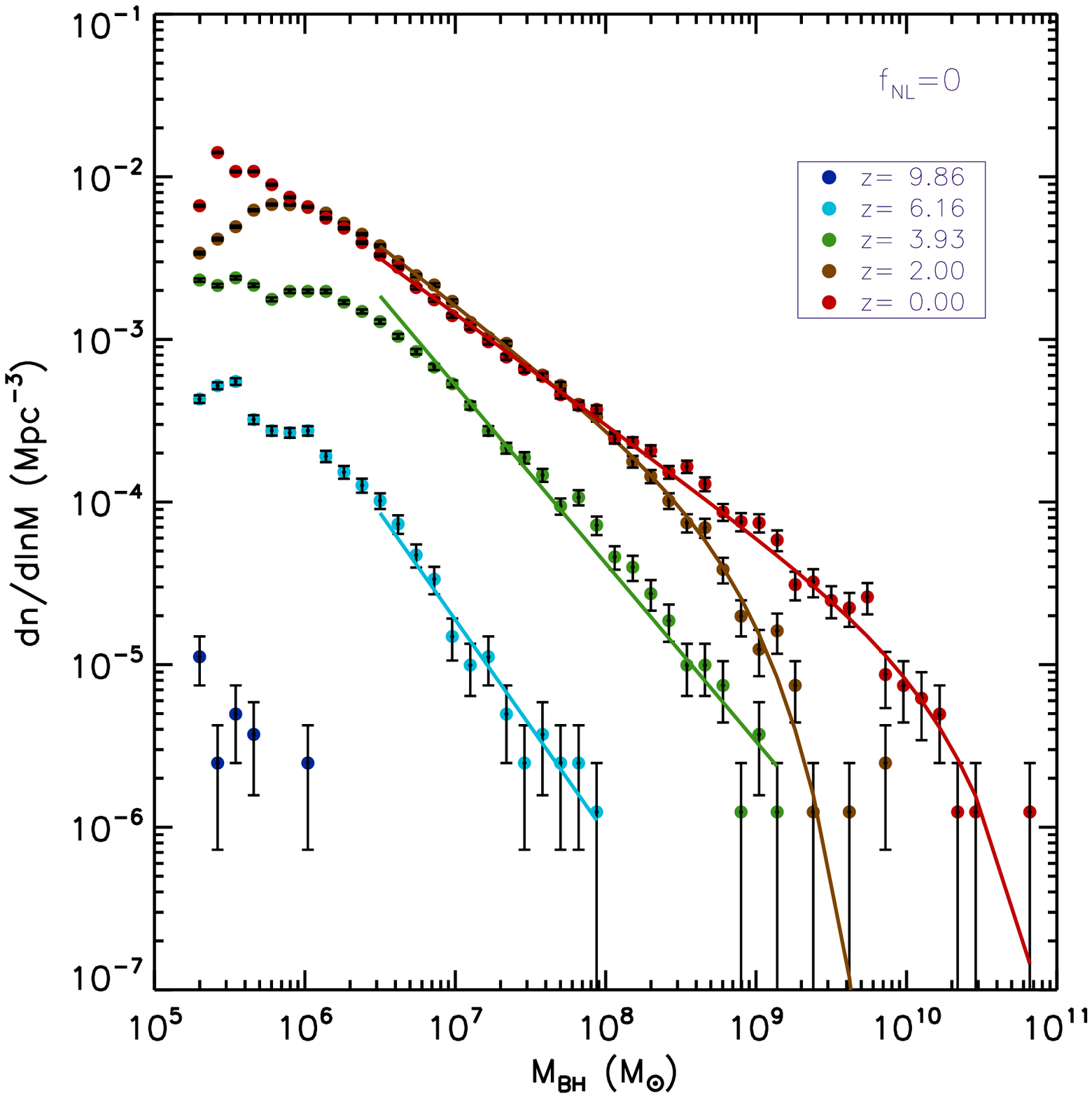}
\includegraphics[trim=0cm 0cm 0cm 0cm, clip=true, angle=0, width=2.3in]{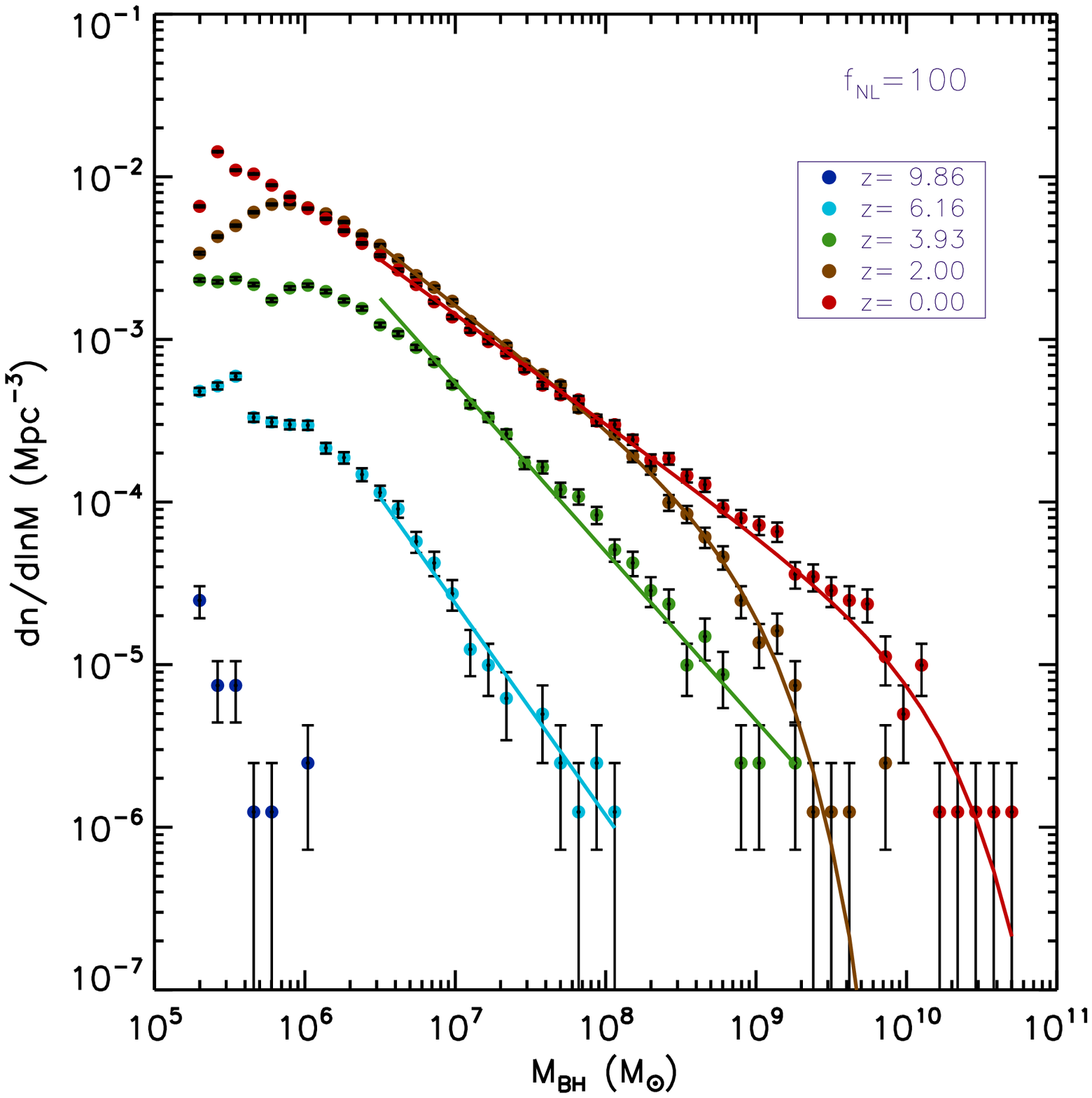}
\includegraphics[trim=0cm 0cm 0cm 0cm, clip=true, angle=0, width=2.3in]{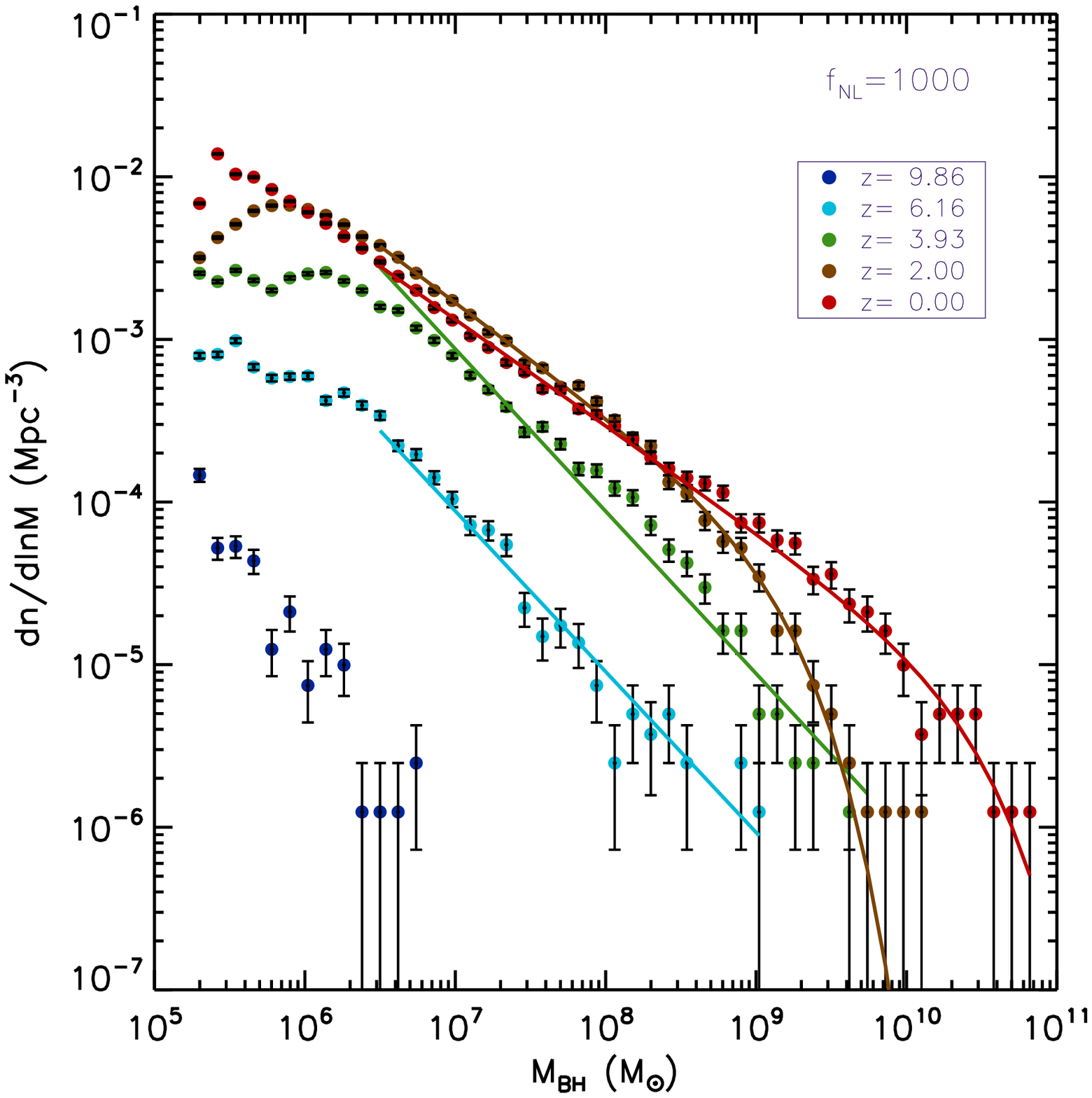}
\caption{The black hole mass function at various redshifts from simulations with $f_{NL}=0, 100, 1000$, respectively. A Schechter function is used as the fitting curve to the simulation data at low redshifts ($z <4$), while a power law is used for high-redshift data ($z >4$). }
\label{fg_bh_mass_func}
\end{center}
\end{figure*}

\begin{figure}
\begin{center}
\includegraphics[angle=0, width=3.4in]{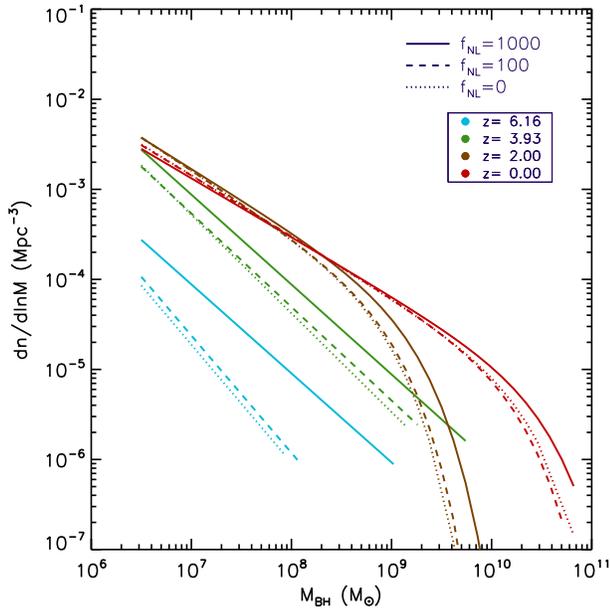}
\caption{A comparison of the fitting curves of the black hole mass functions from all $f_{NL}=0, 100, 1000$ simulations as shown in Fig. \ref{fg_bh_mass_func}.}
\label{fg_bh_mass_func_comp}
\end{center}
\end{figure}

To show the impacts of the primordial NG on the black hole population we plot the black hole mass function of the Gaussian ($f_{NL}=0$) and the non-Gaussian ($f_{NL}=100, 1000$) simulations at various redshifts in Fig. \ref{fg_bh_mass_func}. The black hole mass in this plot corresponds to the mass of every individual black hole in the simulations. Similar to the method used for the stellar mass function in Fig. \ref{fg_stellar_mass_func}, we fit the black hole mass function with a Schechter function (Eq. \ref{eq_schechter_func}) for $z=0$ and $z=2$. For higher redshifts, it doesn't fit a Schechter function well, so we do a simple linear fitting instead. Again, for the data fitting, we only include the black holes with masses at least 20 times of the minimum mass ($10^5\Msun$/h) and don't include $z=9.86$ data. We overplot the fitted curves in Fig. \ref{fg_bh_mass_func} and replot the fitted curves in all $f_{NL}=0, 100, 1000$ simulations in Fig. \ref{fg_bh_mass_func_comp} for a clear comparison.

As shown in Fig. \ref{fg_bh_mass_func} and Fig. \ref{fg_bh_mass_func_comp}, the black hole mass function of the non-Gaussian, $f_{NL}=100$, simulation has a slightly higher number density than the Gaussian one at all redshifts plotted, except at $z=0$ where the trend reverses. We think the reverse trend at $z=0$ is probably due to the limited data in the fitting and thus is not statistically significant. Similar to the trend for the stellar mass function, the excess of the number density is bigger at high redshifts and for more luminous galaxies and smaller at low redshifts and for less luminous galaxies. The $f_{NL}=1000$ case shows a similar trend as for the $f_{NL}=100$ case but with a much bigger excess of the number density especially at high redshifts. The excess increases with redshift and close to zero at the low luminous ends at low redshifts. The difference between the Gaussian and non-Gaussian black hole mass functions is closely related to the cosmic black hole accretion rate shown in Fig. \ref{fg_bhar}. To further investigate the impacts of the primordial NG on the formation and evolution of the black holes, we will study the growth history of individual black holes next.

\begin{figure*}
\begin{center}
\includegraphics[trim=0cm 0cm 0cm 0cm, clip=true, angle=0, width=2.3in]{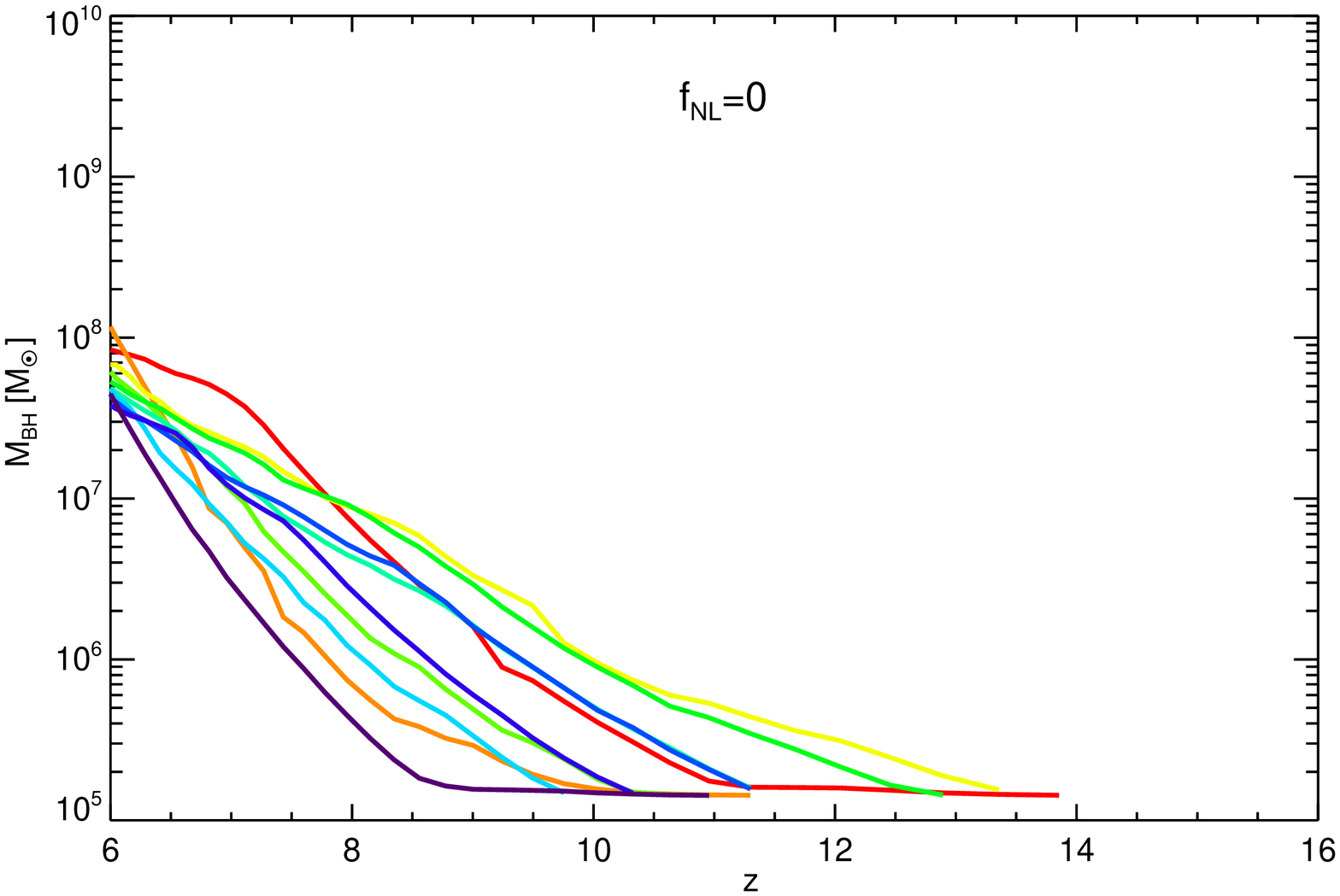}
\includegraphics[trim=0cm 0cm 0cm 0cm, clip=true, angle=0, width=2.3in]{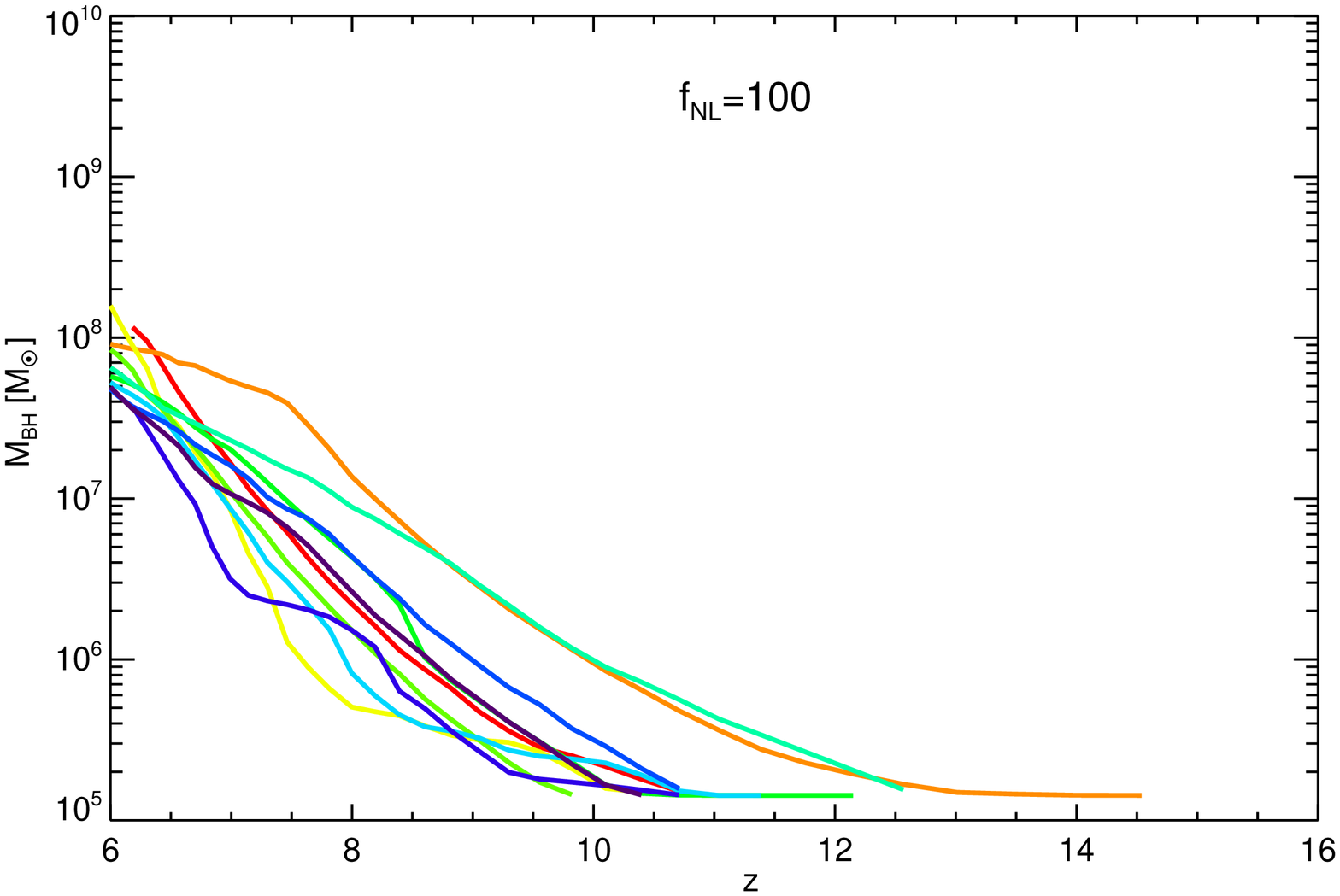}
\includegraphics[trim=0cm 0cm 0cm 0cm, clip=true, angle=0, width=2.3in]{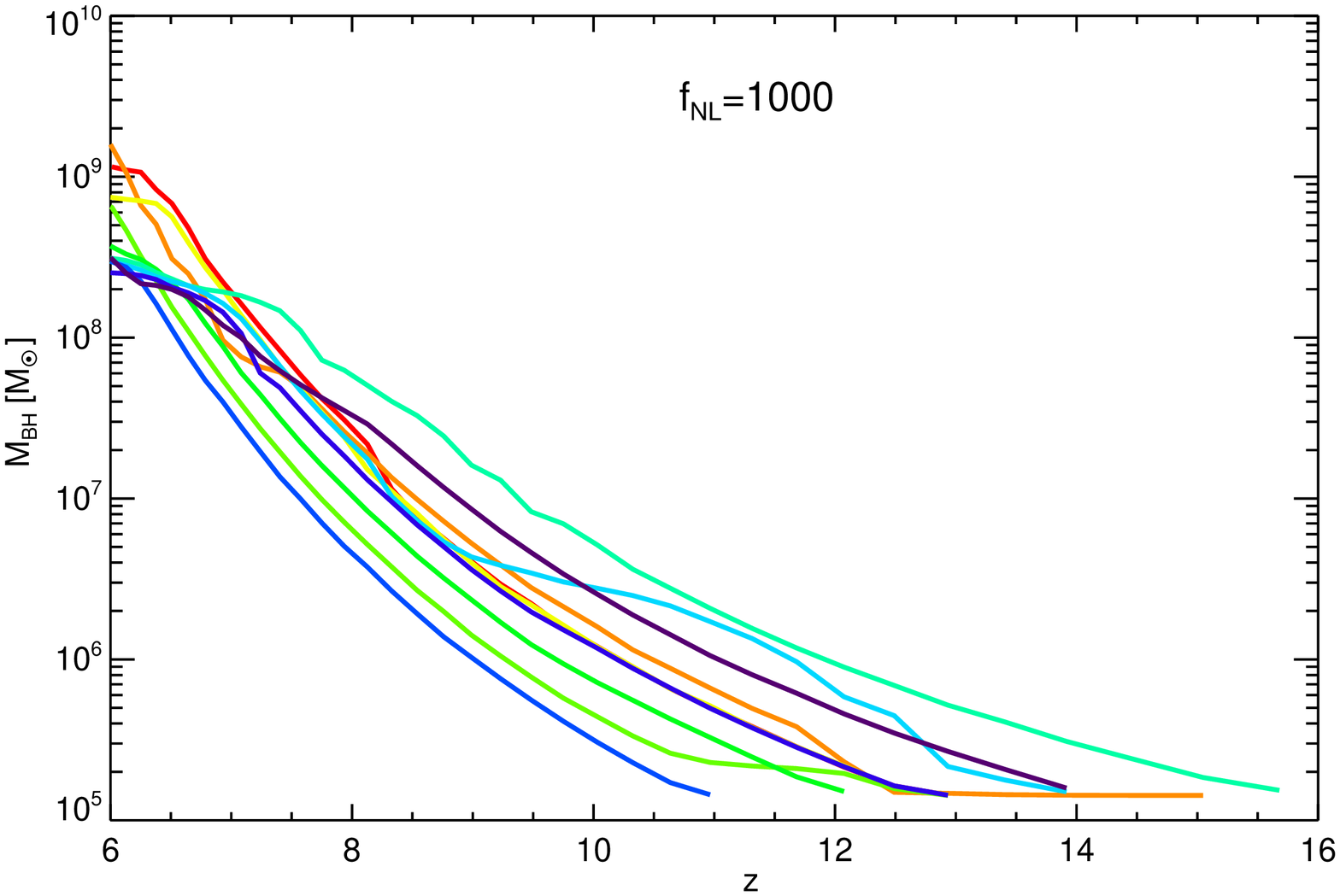} \\
\includegraphics[trim=0cm 0cm 0cm 0cm, clip=true, angle=0, width=2.3in]{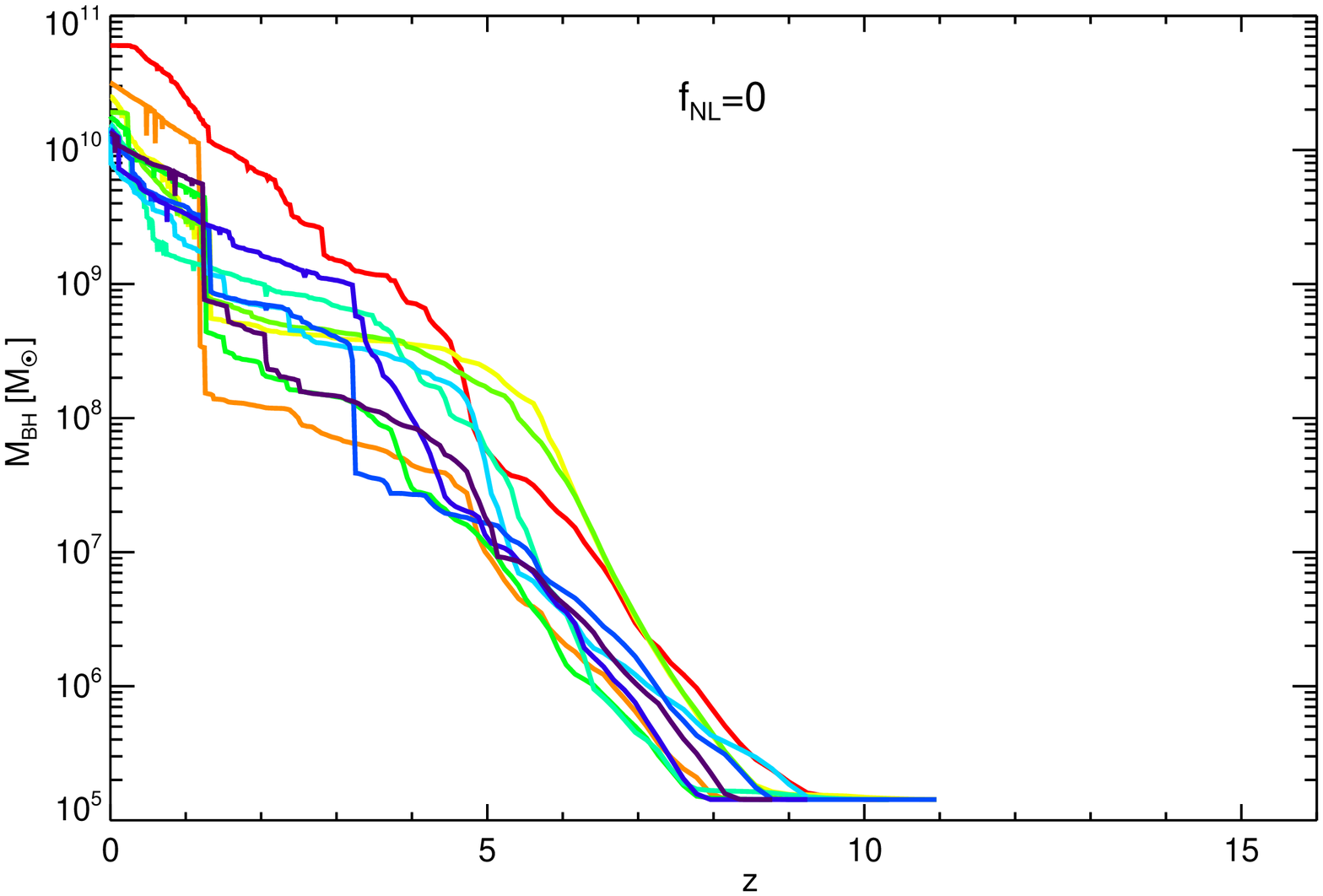}
\includegraphics[trim=0cm 0cm 0cm 0cm, clip=true, angle=0, width=2.3in]{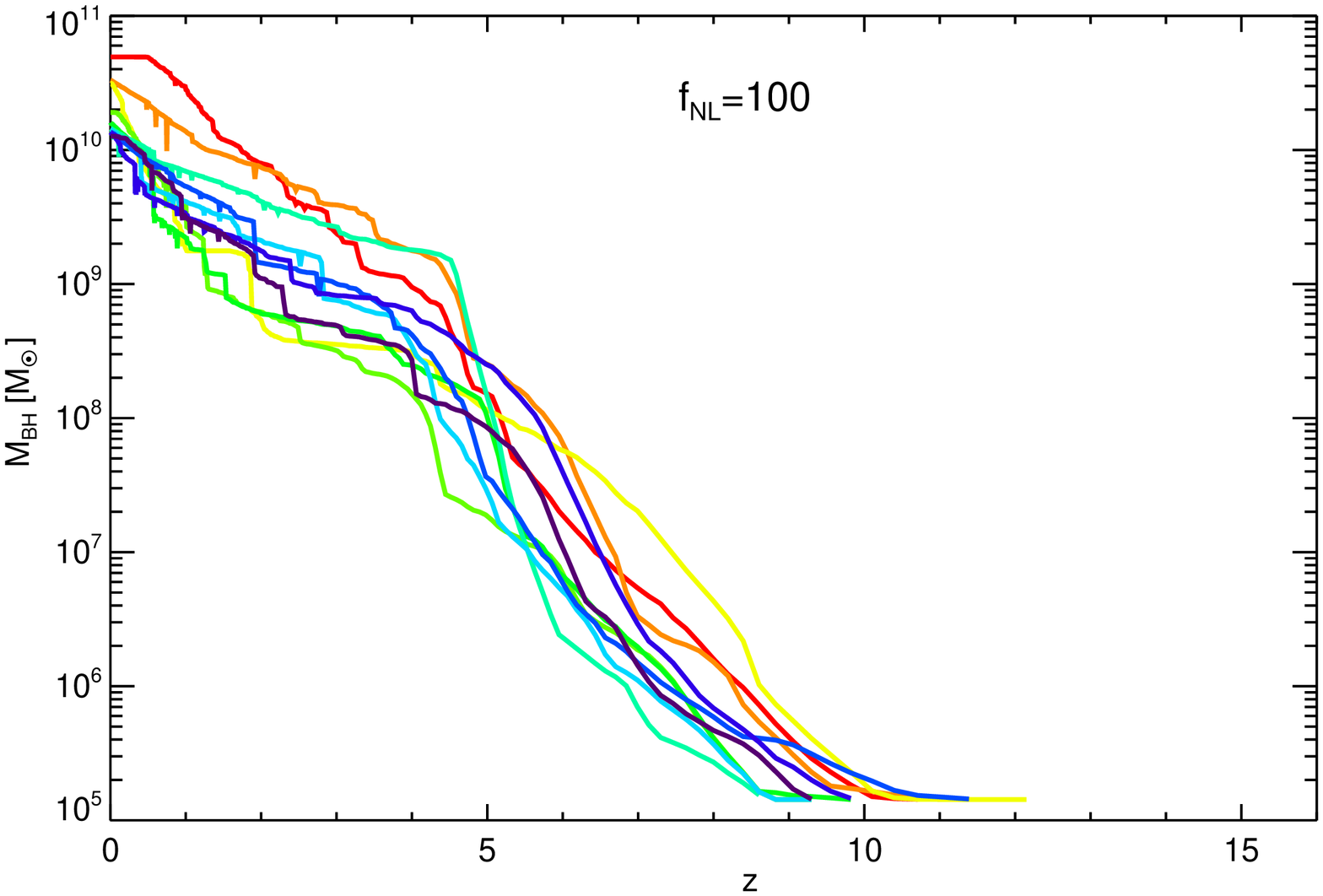}
\includegraphics[trim=0cm 0cm 0cm 0cm, clip=true, angle=0, width=2.3in]{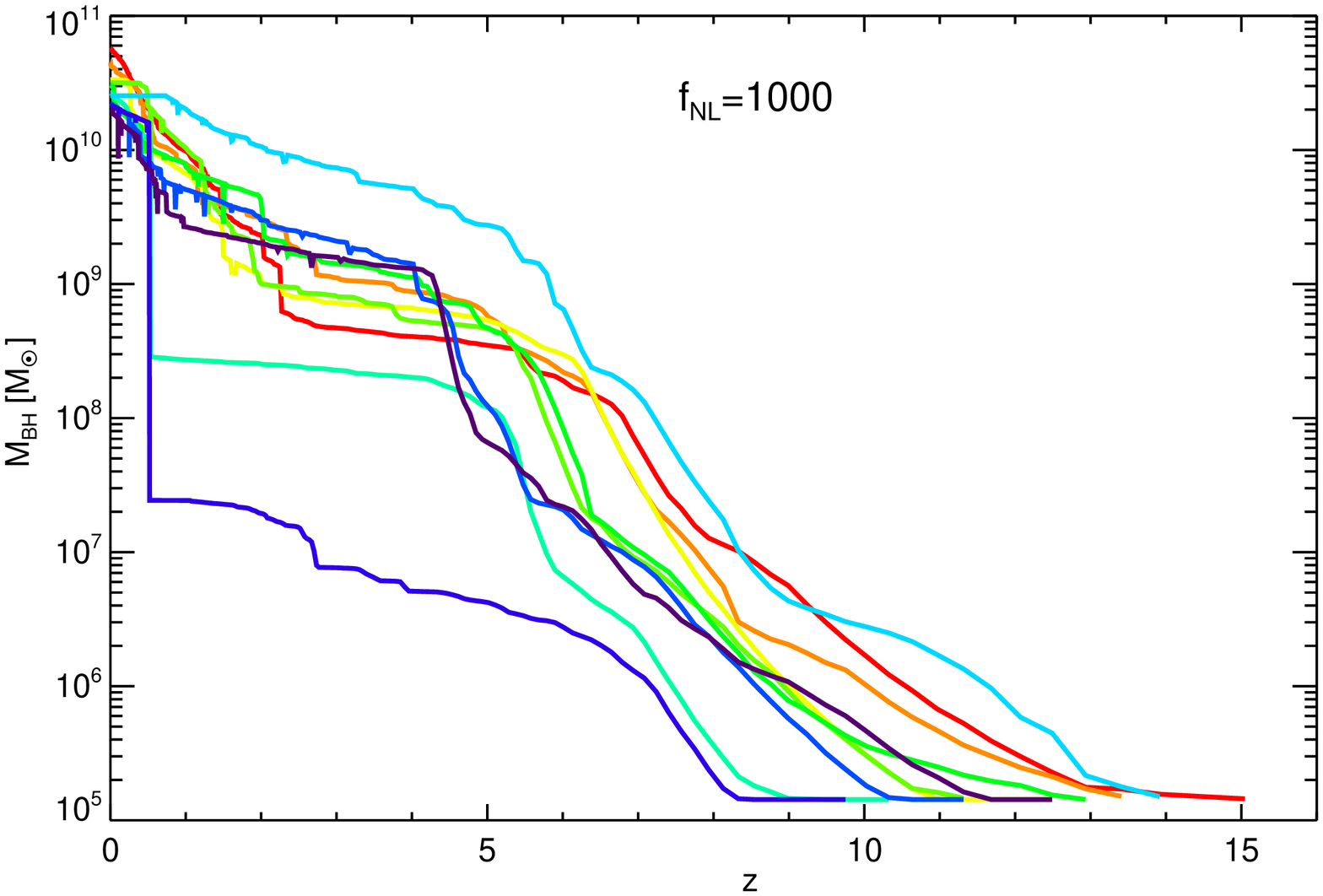}
\caption{A comparison of the growth history of the top 10 most massive black holes at two redshifts, $z=6$ (top panels) and $z=0$ (bottom panels), for simulations with $f_{NL}$=0 (left column), 100 (middle column), and 1000 (right column), respectively. Different color indicates different individual black hole.  
}
\label{fg_bh_growth}
\end{center}
\end{figure*}

The observational data from quasars have offered us some strictest constraints on the primordial NG \citep{Slosar:2008,Xia:2011}. Since quasars are closely related to black hole activities, to improve current methods on constraining the primordial NG using quasars, it is crucial to study black hole growth histories in realistic Gaussian and non-Gaussian simulations. The self-regulated black hole growth model used in this study has been shown to produce results that agree with many observed properties of the local galaxies and distant quasars \citep{Springel:2005b,Springel:2005a,Di-Matteo:2005,Di-Matteo:2008,Hopkins:2006,Li:2007,Zhu:2012}. This enables us to study the black holes in the simulations in great detail. In Fig. \ref{fg_bh_growth}, we plot the growth history of the top 10 black holes in mass at redshift $z\sim6$ and $z=0$ in our Gaussian and non-Gaussian simulations. As we can clearly see from Fig. \ref{fg_bh_growth}, the black holes in the $f_{NL}=1000$ case grow much faster than those in the other two cases at high redshifts. In fact, at redshift $z\sim6$, there have been many black holes with masses around $10^{9}\Msun$ that can well be the central engines of quasars although since our simulations only have a box size of 100 Mpc/h, the number density of quasars is too high at this redshift in the $f_{NL}=1000$ case comparing to the observational data. The difference in the growth history of  the top 10 black holes between the $f_{NL}=100$ and $f_{NL}=0$ cases is not substantial at $z\sim6$. For redshift $z=0$, the final top 10 black hole masses are similar in all three cases although the higher the $f_{NL}$ is, the faster the accretion is at high redshifts, the earlier a black hole's mass reaches $10^{9}\Msun$ with a slower accretion afterwards in general.

\subsection{Galaxy Bias}
\label{ss34}

The scale-dependent bias of galaxies and quasars from large-scale structure surveys has provided constraints on the primordial NG comparable to those from the CMB data \citep{Slosar:2008,Xia:2011,Ross:2013}. Furthermore, it has the potential to achieve a much tighter constraint  ($\mathcal{O}(1)$) using future observations \citep{Carbone:2008,Carbone:2010,Giannantonio:2012,Amendola:2012}. However, most of the theoretical models of bias have been based on N-body simulations with only dark matter (e.g., \citealt{Dalal:2008, Giannantonio:2010, Wagner:2010, Shandera:2011}). Since both the formation and evolution of the visible cosmic structures involve complex baryonic physics, it is imperative to include baryons and the relevant processes in the simulations. In this subsection, we will focus on the differences between the baryonic and dark matter-only simulations, and their possible implications on the results from large-scale structure surveys.

To study the spatial distribution of matter in the simulations, we first calculate the matter power spectrum $P_{mm}(k)$, which describes the density fluctuation as a function of scale in $k$ space. We calculate the matter density fluctuation, $\delta_m(r)$, on a spatial grid of $1024^3$ cells in real space, then convert it to $k$ space, $\hat\delta_m(k)$, using the Fourier transformation. The power spectrum is calculated as follows: 

\begin{equation}
P_{mm}(k)=\langle|\hat\delta_m(k)|^2\rangle
\end{equation}
where the brackets represent the average of all $k$-modes in a particular $k$ bin. For consistency, $P_{mm}(k)$ of the baryonic simulations is calculated only from the dark matter particle distribution.

\begin{figure}
\begin{center}
\includegraphics[angle=0, width=3.4in]{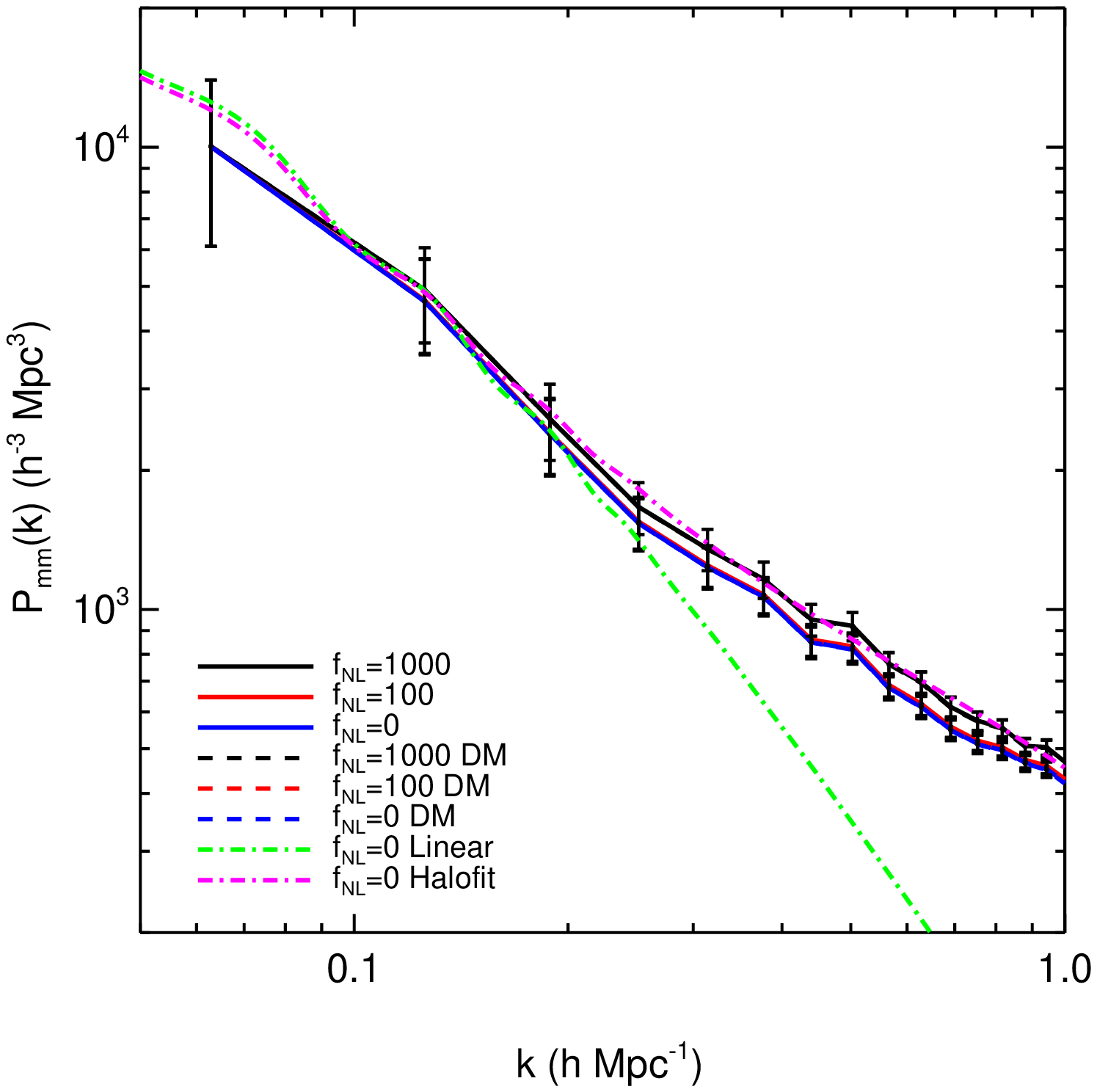} \\
\includegraphics[angle=0, width=3.4in]{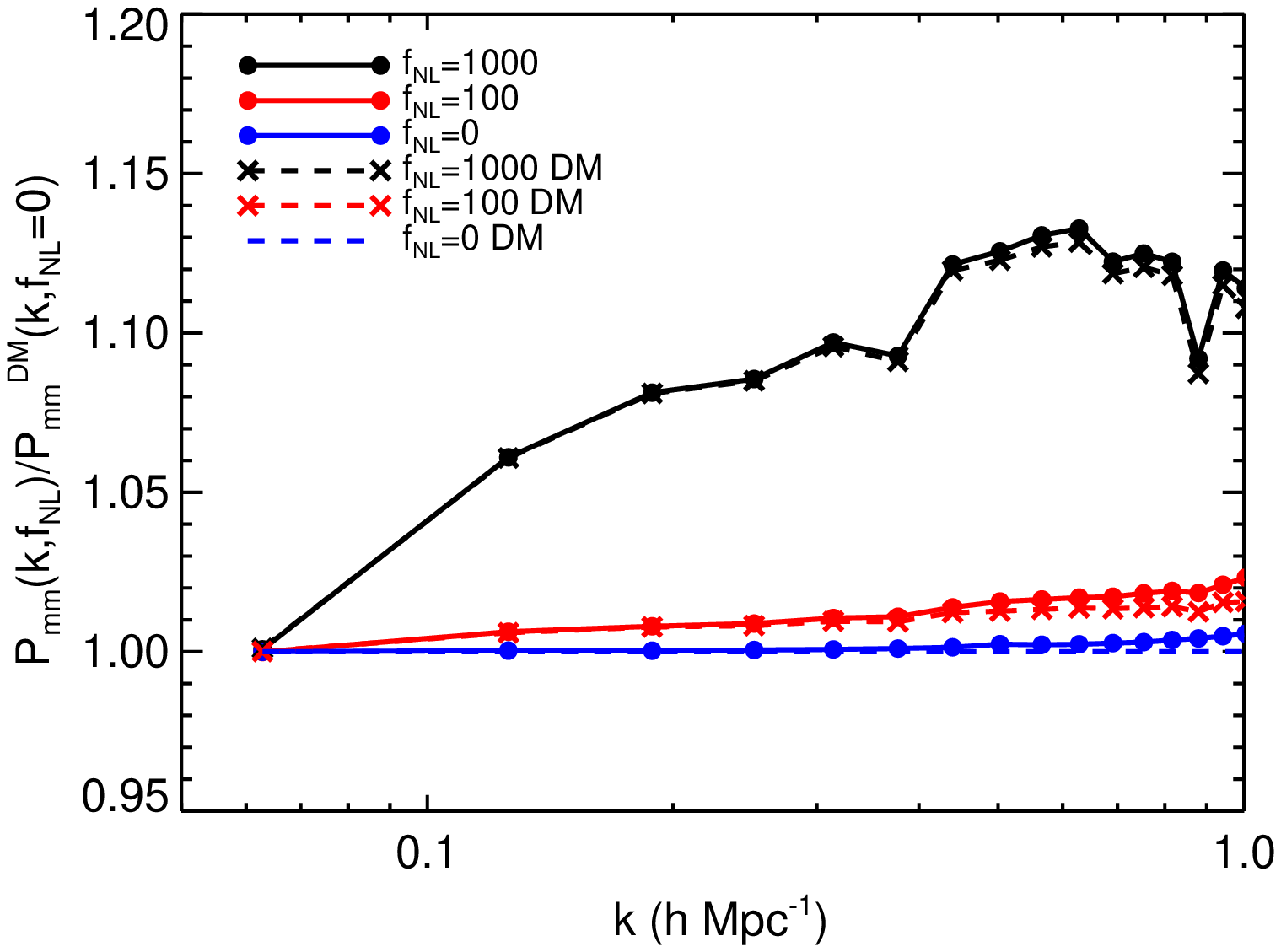}
\caption{Effects of baryon and NG on the matter power spectrum. The top panel shows the $P_{mm}(k)$ of both baryonic and dark matter-only simulations at $z=0$ for $f_{NL}=0, 100, 1000$, respectively, in comparison with the linear theory and the Halofit model \citep{Smith:2003}, as indicated in the legend. The bottom panel shows the relative ratio of $P_{mm}(k, f_{NL})$ to the dark matter-only, $f_{NL}=0$ simulation, $P_{mm}(k, f_{NL})/P_{mm}^{DM}(k, f_{NL}=0)$.}
\label{fg_power_spec_comp}
\end{center}
\end{figure}

To investigate the effects of baryon and NG on the matter power spectrum, we show in Figure~\ref{fg_power_spec_comp} the resulting $P_{mm}(k)$ from both baryonic and dark matter-only simulations, each includes three different initial conditions with $f_{NL}=0, 100, 1000$, in comparison with the Halofit model \citep{Smith:2003} and the linear theory (top panel), as well as the relative ratio of $P_{mm}(k, f_{NL})$ to the dark matter-only, $f_{NL}=0$ simulation (bottom panel). 

As we can see from Fig. \ref{fg_power_spec_comp}, the $P_{mm}(k)$ from all six simulations agree with each other very well within their error bars which are calculated from $P_{mm}(k)/\sqrt{N_k}$ where $N_k$ is the number of independent $k$-modes per $k$ bin. $P_{mm}(k)$ from the simulations all agree with the prediction from the Halofit model, too. In order to see the differences in $P_{mm}(k)$ among the simulations, in the bottom panel of Fig. \ref{fg_power_spec_comp}, we plot their relative ratios to the dark matter only, $f_{NL}=0$ simulation. As we can see, in general, $P_{mm}(k)$ increases with $f_{NL}$, so matter in the non-Gaussian cases is more clustered than that in the Gaussian cases. It is also shown that on the large scale ($k\le1.0$ h/Mpc), $P_{mm}(k)$ from the baryonic and dark matter only simulations roughly agree with each other. We also compare our results from dark matter only simulations with Fig. 3 in \cite{Giannantonio:2010} and Fig. 4 in \cite{Wagner:2010}, our results show good agreements with the ones in these studies as predicted in \cite{Taruya:2008}.

\begin{figure}
\begin{center}
\includegraphics[angle=0, width=3.4in]{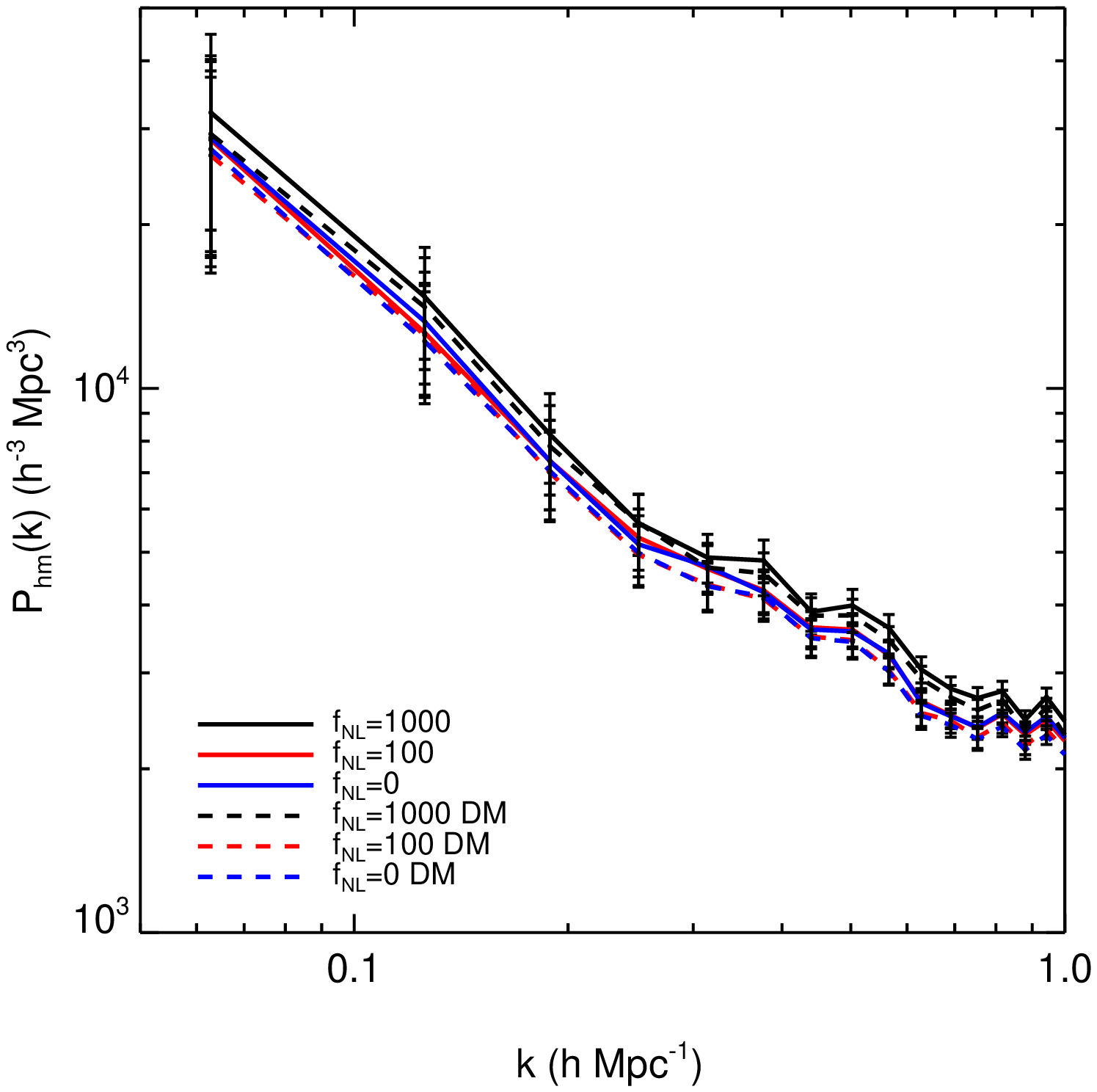} \\
\includegraphics[angle=0, width=3.4in]{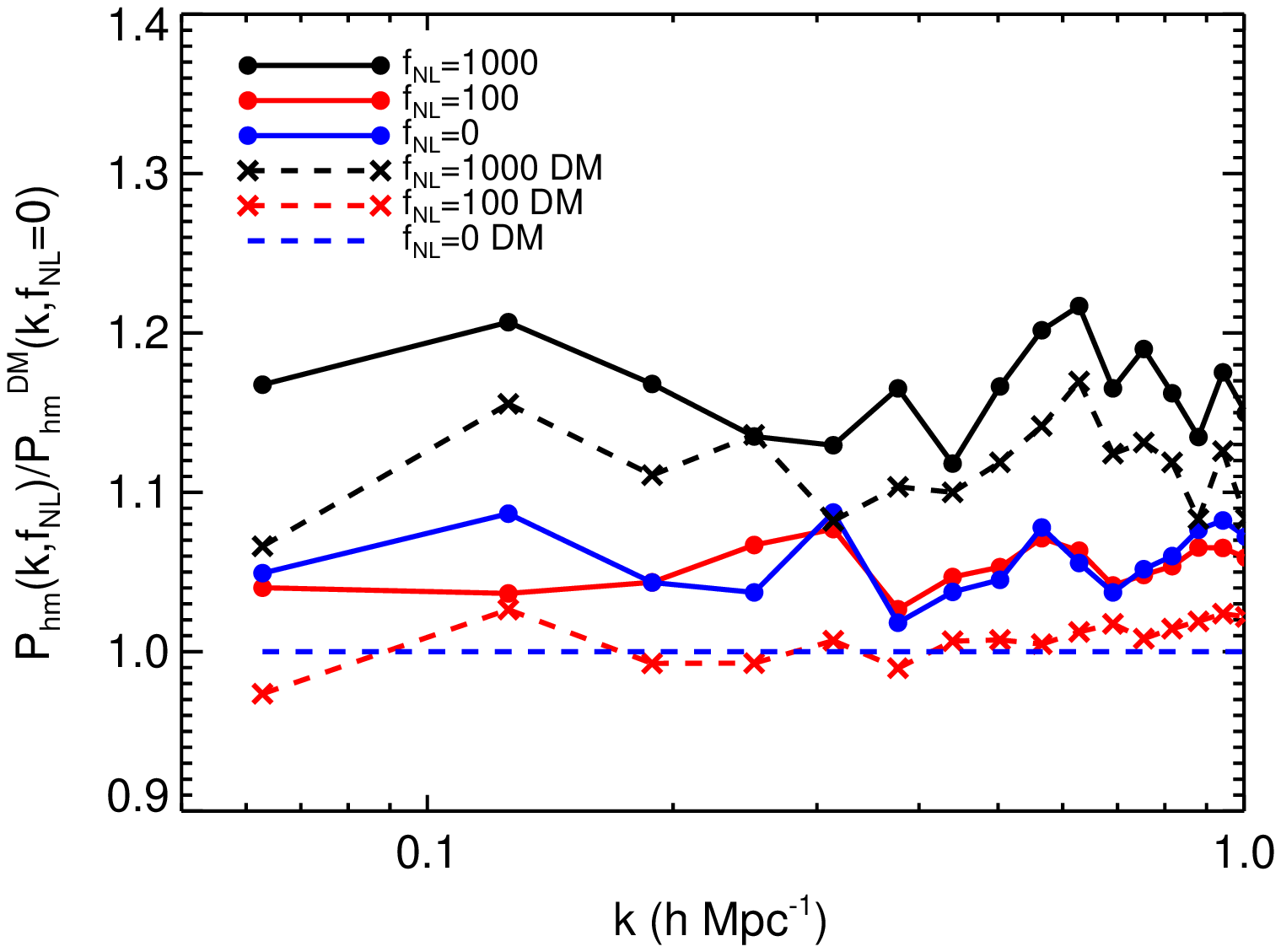}
\caption{Effects of baryon and NG on the halo-matter cross spectrum. The top panel shows the $P_{hm}(k)$ of both baryonic and dark matter-only simulations at $z=0$ for $f_{NL}=0, 100, 1000$, respectively. The bottom panel shows the relative ratio of $P_{hm}(k, f_{NL})$ to the dark matter-only, $f_{NL}=0$ simulation, $P_{hm}(k, f_{NL})/P_{hm}^{DM}(k, f_{NL}=0)$.}
\label{fg_cross_spec_comp}
\end{center}
\end{figure}

We now study the clustering properties of the galaxies (halos) in the simulations. In the top panel of Fig. \ref{fg_cross_spec_comp}, we plot the galaxy (halo) matter cross spectrum, $P_{hm}(k)$, of the baryonic and dark matter only simulations for $f_{NL}=0, 100, 1000$ cases. We calculate $P_{hm}(k)$ in a way similar to $P_{mm}(k)$ as $P_{hm}(k)=\langle {\rm{Re}}(\hat\delta_h(k)\hat\delta_m^\ast(k)) \rangle$, where $\hat\delta_h(k)$ is the halo density fluctuation in $k$ space. For an easy comparison, we will treat both galaxies and halos as same type of objects for the study of clustering from now on. As shown in Fig. \ref{fg_cross_spec_comp}, $P_{hm}(k)$ from all six simulations again agree with each other within their error bars. To see the differences in $P_{hm}(k)$ among the simulations, in the bottom panel of Fig. \ref{fg_cross_spec_comp}, we plot their relative ratios to the dark matter only, $f_{NL}=0$ simulation. As we can see, $P_{hm}(k)$ in the baryonic simulations is bigger than that in the dark matter only simulations on the full scale plotted for same $f_{NL}$. So galaxies in the baryonic simulations are more clustered than the halos in the dark matter only simulations. Since the bias is defined as: $b=P_{hm}/P_{mm}$ and $P_{mm}$ is almost same for both types of simulations for $k\le1.0$ h/Mpc in this study, the baryonic simulations will have bigger bias than the dark matter only simulations on the large scale as we will show later in this section. For redshift $z=0$, the difference in $P_{hm}(k)$ between the $f_{NL}=100$ and $f_{NL}=0$ cases is very small for both types of simulations and in order to illustrate the baryonic effects on the bias efficiently, we will mainly focus on the $f_{NL}=1000$ case in the analysis of the bias in this section.

\begin{figure}
\begin{center}
\includegraphics[angle=0, width=3.4in]{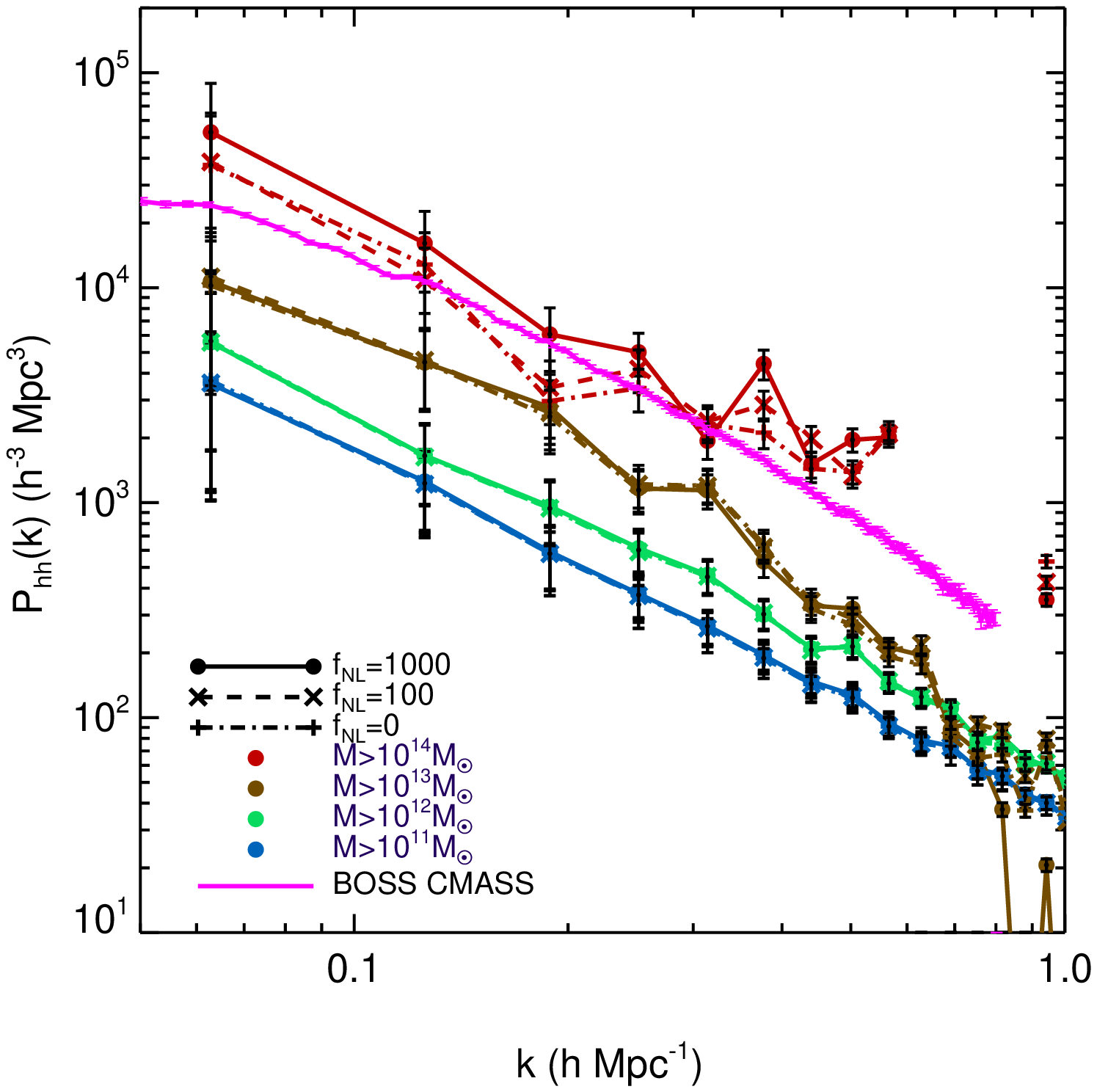}
\caption{Halo power spectrum in different mass range from the baryonic simulations with $f_{NL}=0, 100, 1000$, respectively, in comparison with that from the BOSS CMASS sample \citep{Anderson:2012, Ross:2013}. Different color indicates different halo mass range.}
\label{fg_group_spec_comp}
\end{center}
\end{figure}

Since the bias generally depends on the mass of the objects that are used as tracers of the matter density field, we want to choose a suitable mass range for the halos to study the bias in the simulations. In Fig. \ref{fg_group_spec_comp}, we plot the halo power spectrum of the baryonic simulations for $f_{NL}=0, 100, 1000$ cases at different mass range and compare it to the CMASS sample from the SDSS DR9 BOSS data \citep{Anderson:2012, Ross:2013}. The shot noise ($1/n_h$ where $n_h$ is the halo number density) has been subtracted from the simulation data and the simulation data have been scaled to CMASS sample redshift $z=0.57$ using the growth function. As shown in Fig. \ref{fg_group_spec_comp}, the halo group with mass bigger than $10^{14} \Msun$ will have a bias close to that of the CMASS sample. We will use this halo group for the analysis of the bias next.

\begin{figure}
\begin{center}
\includegraphics[angle=0, width=3.4in]{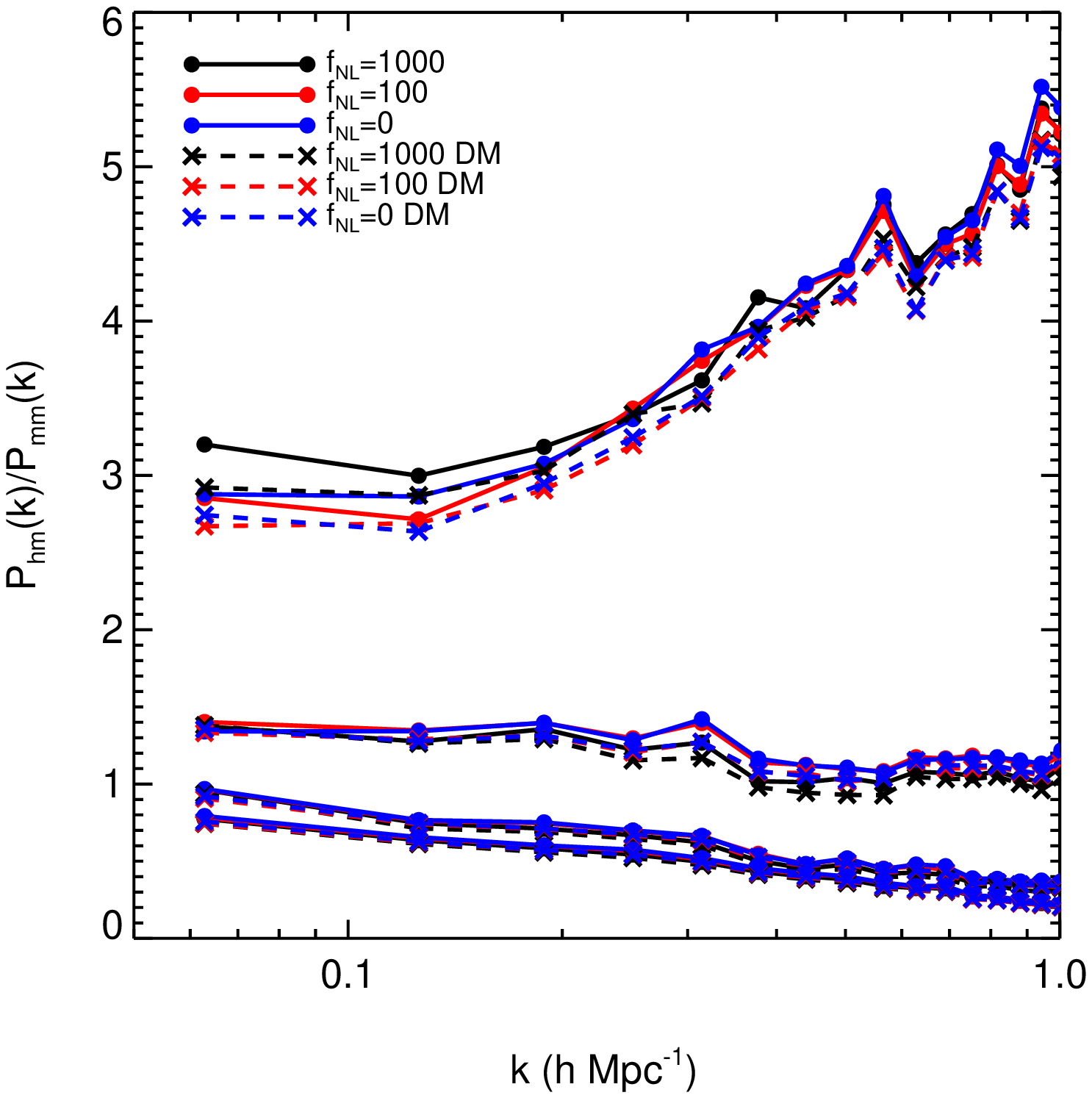}
\caption{Halo bias for halos more massive than $10^{14} \Msun$, $10^{13} \Msun$, $10^{12} \Msun$ and $10^{11} \Msun$ (from top to bottom) in the baryonic (solid lines) and dark matter-only (dashed lines) simulations for $f_{NL}=0, 100, 1000$, respectively. Different color indicates different $f_{NL}$.}
\label{fg_bias_gas_dm}
\end{center}
\end{figure}

In order to study the difference in the halo bias between the baryonic and dark matter only simulations, in Fig. \ref{fg_bias_gas_dm}, we plot the halo bias calculated from both types of simulations for $f_{NL}=0, 100, 1000$ cases for halos more massive than $10^{14} \Msun$. As we expect from the analyses in the previous paragraphs, the galaxies in the baryonic simulations have bigger bias than the halos in the dark matter only simulations. For the $f_{NL}=1000$ cases, both types of simulations show larger bias than the $f_{NL}=0$ cases on the large scale and show the trend that the bias will increase with the scale on the largest scale in the simulations. Again, there is no clear trend shown between $f_{NL}=100$ and $f_{NL}=0$ cases due to the small box size of our simulations.

\begin{figure}
\begin{center}
\includegraphics[angle=0, width=3.4in]{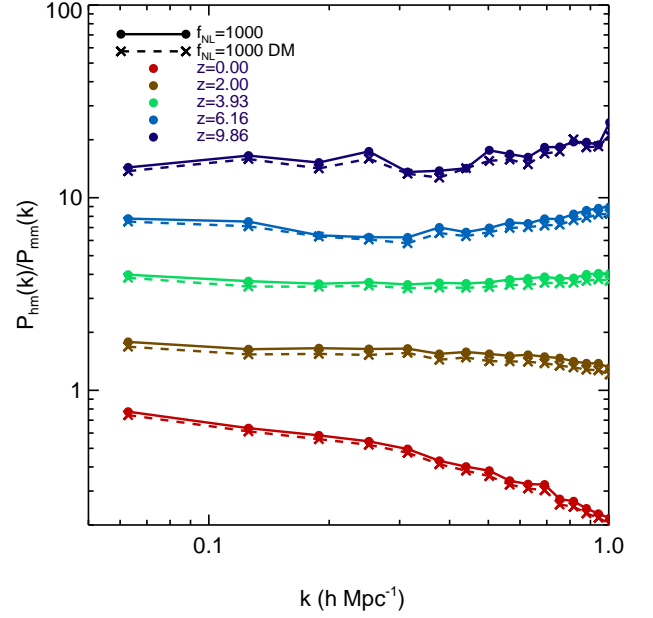}
\caption{Halo bias for halos more massive than $10^{11} \Msun$ at various redshifts from both the baryonic (solid lines) and dark matter-only (dashed lines) simulations with $f_{NL}=1000$. Different color indicates different redshift.}
\label{fg_bias_z_gas_dm}
\end{center}
\end{figure}

Now we want to study the time evolution of the halo bias from the baryonic and dark matter only simulations. In Fig. \ref{fg_bias_z_gas_dm}, we plot the halo bias for selected halo group in both baryonic and dark matter only simulations for $f_{NL}=1000$ case at five redshifts. It is clearly shown that the galaxies in the baryonic simulations always have bigger bias than the halos in the dark matter only simulations in all redshifts plotted. The halo bias on the largest scale in the simulations doesn't always increase with the scale as expected, we think this is due to the small box size of our simulations so that the structures are suppressed on the scale close to the box size.

\section{Discussions and Conclusions}
\label{s4}

In this study, we have used cosmological hydrodynamical simulations that include the baryonic components and corresponding physics to investigate the effects of the primordial NG on the structure formation and evolution in the Universe. We have confirmed some previous results on the properties of the baryonic components in the non-Gaussian numerical simulations and expanded our investigations to new physical quantities. We have compared our results from the simulations with the theoretical predictions and the observational data. Our main findings on the structures on different scales are as follows: 

\begin{enumerate}

\item For the galaxy mass function, the MVJ and LMSV formulations are still roughly good for the simulations with baryonic components. 

\item The primordial NG has a substantial impacts on the cosmic star formation rate density at high redshifts. The star formation rate density may be used to significantly constrain the primordial NG if the star formation rate in the simulation is well calibrated and high redshift star formation rate data from observations become available. 

\item The black hole growth history is very sensitive to the magnitude of the primordial NG. Thus high redshift quasars are useful tools to constrain the primordial NG. 

\item The primordial NG has substantial impacts on both the stellar and black hole mass functions especially at high redshifts. 

\end{enumerate}

However, we should point out that this study has suffered from a number of limitations due to our poor understanding of physical processes in galaxy formation and the cost of computations : 

\begin{itemize}

\item Small box size. Because it's extremely expensive to perform cosmological simulations with baryons and related physical processes in a large volume, all the simulations in this work were performed in a box with a side length of 100 Mpc. This small volume limits our ability to investigate  structure properties in small k (i.e., $k < 0.01$),  and to compare our results with those from large-box ($\sim$ Gpc scale), DM-only simulations.  

\item Small number of realizations. In this work, we have performed two sets of simulations, baryonic and DM-only, each with $f_{NL}=0, 100, 1000$, respectively. For each of the case, only one realization was done. More realizations are needed for a systematic study in order to constrain the dispersion of the properties we study here.

\item Simplified sub-grid models for baryon physics. In large-scale cosmological simulations like the ones we perform here, it is impossible to follow star formation and black hole growth from first principles due to limited resolutions. We have used sub-grid recipes for a number of physical processes including star formation, black hole accretion, and feedback. While the star formation model generally reproduces the cosmic star formation history, the black hole model appears to overpredict massive black holes, in particular those with mass $\sim 10^{11}\, \Msun$, which have not been found in the local Universe.   

\end{itemize}

While these limitations would affect the detailed properties such as mass functions of galaxies and black holes, they should not affect our results on the {\it relative effects} of non-Gaussianity on structure formation, which are the focus of this paper. We plan to improve this study by performing a systematic study (more realizations) in a larger volume, and with an improved black hole model which reproduce the BH mass functions and the cosmic accretion history, in future work. 

\section*{Acknowledgments}

Support from NSF grants AST-0965694 and AST-1009867 is gratefully acknowledged. SS acknowledges support from NASA NNX12AC99G. DJ acknowledges the support of DoE SC-0008108 and NASA NNX12AE86G. We acknowledge the Research Computing and Cyberinfrastructure unit of Information Technology Services at the Pennsylvania State University for providing computational resources and services that have contributed to the research results reported in this paper (URL: http://rcc.its.psu.edu). The Institute for Gravitation and the Cosmos is supported by the Eberly College of Science and the Office of the Senior Vice President for Research at the Pennsylvania State University.

\bibliography{non_gaussianity}

\end{document}